\documentclass[twocolumn,preprint]{aastex631}

\usepackage{soul}
\usepackage{color}

\usepackage{CJK}
\usepackage{apjfonts}
\usepackage{amsmath}
\usepackage{color}

\hypersetup{
   colorlinks,
   linkcolor={blue!88!black!80},
   citecolor={blue!88!black!80},
   urlcolor={blue!88!black!80}}

\shorttitle{IMBH INOV}
\shortauthors{Zuo et al.}

\turnoffeditone

\begin{document}
\begin{CJK}{UTF8}{gbsn}

\title{Constraints on the Intranight Optical Variability of Intermediate-Mass Black Hole Candidates}

\correspondingauthor{Wenwen Zuo}
\email{wenwenzuo@shao.ac.cn (WWZ)}
\email{hengxiaoguo@gmail.com (HXG)}

\author[0000-0002-4521-6281]{Wenwen Zuo}
\affiliation{Shanghai Astronomical Observatory, Chinese Academy of Sciences, 80 Nandan Road, Shanghai 200030, People's Republic of China}

\author[0000-0001-8416-7059]{Hengxiao Guo} 
\affiliation{Shanghai Astronomical Observatory, Chinese Academy of Sciences, 80 Nandan Road, Shanghai 200030, People's Republic of China}

\author{Wanling Liu} 
\affiliation{Shanghai Astronomical Observatory, Chinese Academy of Sciences, 80 Nandan Road, Shanghai 200030, People's Republic of China}
\affiliation{University of Chinese Academy of Sciences, 19A Yuquan Road, 100049, Beijing, People's Republic of China}

\author[0000-0002-3742-6609]{Wenke Ren} 
\affiliation{Shanghai Astronomical Observatory, Chinese Academy of Sciences, 80 Nandan Road, Shanghai 200030, People's Republic of China}

\author[0000-0001-8416-7059]{Jingbo Sun} 
\affiliation{Shanghai Astronomical Observatory, Chinese Academy of Sciences, 80 Nandan Road, Shanghai 200030, People's Republic of China}
\affiliation{University of Chinese Academy of Sciences, 19A Yuquan Road, 100049, Beijing, People's Republic of China}

\author[0000-0001-7566-7561]{Patricia Ar\'{e}valo}
\affiliation{1 Instituto de F\'{i}sica y Astronom\'{i}a, Facultad de Ciencias, Universidad de Valpara\'{i}so, Gran Bretaña 1111, Valpara\'{i}so, Chile}
\affiliation{2 Millennium Nucleus on Transversal Research and Technology to Explore Supermassive Black Holes (TITANS)}

\author[0000-0001-6947-5846]{Luis C. Ho}
\affiliation{Department of Astronomy, School of Physics, Peking University, Beijing 100871, People's Republic of China} 
\affiliation{Kavli Institute for Astronomy and Astrophysics, Peking University, Beijing, 100871, People's Republic of China} 

\author[0000-0002-9331-4388]{Alok C. Gupta}
\affiliation{Aryabhatta Research Institute of Observational Sciences (ARIES), Manora Peak, Nainital 263001, India} 
\affiliation{Xinjiang Astronomical Observatory, Chinese Academy of Sciences, 150 Science 1-Street, Urumqi 830011, People's Republic of China
}

\author[0000-0001-5878-4811]{Vineet Ojha} 
\affiliation{Kavli Institute for Astronomy and Astrophysics, Peking University, Beijing, 100871, People's Republic of China} 

\author[0000-0002-0771-2153]{Mouyuan Sun}
\affiliation{Department of Astronomy, Xiamen University, Xiamen, Fujian 361005, People's Republic of China} 

\author[0000-0002-7299-4513]{Shuang-liang Li}
\affiliation{Shanghai Astronomical Observatory, Chinese Academy of Sciences, 80 Nandan Road, Shanghai 200030, People's Republic of China}

\author[0000-0002-1530-2680]{Haicheng Feng}
\affiliation{Yunnan Observatories, Chinese Academy of Sciences, 396 Yangfangwang, Guandu District, Kunming 650216, Yunnan, People's Republic of China}
\affiliation{Key Laboratory for the Structure and Evolution of Celestial Objects, Chinese Academy of Sciences, Kunming 650216, Yunnan,
People's Republic of China}
\affiliation{Center for Astronomical Mega-Science, Chinese Academy of Sciences, 20A Datun Road, Chaoyang District, Beijing 100012,
People's Republic of China}

\author[0000-0003-4671-1740]{Qi Yuan}
\affiliation{Changchun Observatory, National Astronomical Observatories, Chinese Academy of Sciences, Changchun 130117, People's Republic of China}

\author[0000-0002-4455-6946]{Minfeng Gu}
\affiliation{Shanghai Astronomical Observatory, Chinese Academy of Sciences, 80 Nandan Road, Shanghai 200030, People's Republic of China}

\author[0000-0002-7350-6913]{Xuebing Wu}
\affiliation{Department of Astronomy, School of Physics, Peking University, Beijing 100871, People's Republic of China} 
\affiliation{Kavli Institute for Astronomy and Astrophysics, Peking University, Beijing, 100871, People's Republic of China} 
\begin{abstract}
Intermediate-mass black holes (IMBHs) provide a unique regime for studying accretion variability at the low-mass end of the black hole population, yet their intranight optical variability (INOV) remains poorly constrained. We present a systematic investigation of INOV in an optically selected sample of IMBH candidates using high-cadence observations from the Zwicky Transient Facility (ZTF). From a parent sample of 1,447 broad H$\alpha$-selected candidates, we identify 64 IMBH candidates (median $f_{\mathrm{AGN}}\sim0.06$) with 163 intranight monitoring sessions. Apparent INOV signals identified by conventional ZTF PSF-fit photometry are largely associated with seeing-dependent changes in the relative contributions of compact nuclear and extended host components, which can mimic intrinsic short-timescale variability. In contrast, no robust INOV is detected with difference-image analysis. An ensemble structure function spanning $\Delta t\sim0.003$--$1600$ days reveals long-term variability in a small subsample of sources, whereas intrinsic variability remains unresolved at intranight timescales. Monte Carlo simulations further show that ZTF-like single-night monitoring has a low INOV recovery probability ($\sim1.2\%$) for the variability amplitudes inferred from the long-term analysis. The recovery probability is primarily controlled by source brightness, AGN contribution, intrinsic variability amplitude, and photometric precision. These results demonstrate that the absence of detected INOV does not imply the absence of rapid accretion variability, but can reflect the limited detectability of low-amplitude signals under current observing capabilities. Our findings highlight the importance of robust photometric methodologies for future high-cadence variability studies of low-mass accreting black holes.
\end{abstract}

\section{Introduction}\label{sec:intro}

Accurate measurements of the masses of intermediate-mass black holes (IMBHs; $M_{\rm BH}\sim10^{2-6}~M_{\odot}$) and a comprehensive census of their population are essential for understanding the formation pathways of BH seeds and the subsequent growth of supermassive black holes (SMBHs) in the early Universe \citep[see reviews, ][]{Mezcua17,Greene2020,Inayoshi20,Volonteri21,Reines22}. Continuum reverberation mapping (RM) provides a unique approach for probing compact accretion disk regions and enabling indirect BH mass measurements through the scaling relation between continuum-emitting disk sizes and broad-line region sizes \citep{Wang2023}. Such measurements also provide important tests of accretion disk models and the physical mechanisms responsible for optical variability in low-mass BH systems \citep{Cackett2021RM}.

Although substantial progress has been made in expanding the population of IMBH candidates \citep{Baldassare2016, MP2020, Burke2020, Treiber2023, Bernal_etal_2025}, a major challenge for variability-based studies is that the observed optical fluctuations of low-mass BH systems can be difficult to detect. The intrinsic variability signal is often diluted by host-galaxy emission, while stochastic flux variations and limited monitoring windows further complicate the detection of short-timescale variability.

NGC 4395 and POX 52 are among the best-studied nearby IMBH host candidates, supported by multi-wavelength evidence including broad H$\alpha$ emission and diagnostic classifications based on the BPT diagram \citep{Filippenko2003, Barth_etal_2004}. NGC 4395, a nearby ($z=0.001$) bulgeless spiral dwarf galaxy hosting the least luminous known Seyfert~1 nucleus, exhibits rapid optical and X-ray variability and has been successfully studied through both broad-line and continuum reverberation mapping (RM) \citep{Filippenko1989, Ho1993, Lira1999, Woo2019, Cho2020, Cho2021, Montano2022, McHardy2023}. In contrast, optical RM constraints for POX~52, a dwarf elliptical galaxy accreting near the Eddington limit \citep{Barth_etal_2004}, remain limited. While broad-line RM has not yet been achieved, recent optical and mid-infrared monitoring detected a mid-infrared lag of $\sim35$ days and only a marginal $g-r$ inter-band lag of $\lesssim1$ day \citep{SunJB_etal_2025}.

Motivated by the successful continuum RM measurements of NGC~4395, \citet{Zuo2024} conducted continuum RM observations of another IMBH candidate, SDSS J024912.86$-$081525.6. Based on the empirical relation between the $g-z$ continuum lag, AGN luminosity at 5100~\AA, and BH mass established from nearby AGNs with multi-band continuum RM measurements, the expected continuum lag was estimated to be only 1.6--3.3~hr. However, no significant variability was detected at the level of $<1.4\%$ over 6--10~hr, likely because host-galaxy contamination diluted the intrinsic AGN variability signal. 

Taken together, these studies show that the detectability of rapid optical variability can differ among individual IMBH candidates. Individual detections or non-detections therefore cannot distinguish intrinsically weak variability from limited observational sensitivity or determine how commonly such variability can be detected across the broader population. This motivates a systematic investigation of intranight optical variability (INOV), i.e., optical flux variations occurring on sub-day timescales within a single monitoring session \citep{Miller_etal_1989,Wagner_Witzel1995}. A statistical assessment of the INOV detection fraction will quantify the practical detectability of short-timescale variability and provide guidance for future high-cadence monitoring campaigns, including target selection and the photometric precision required to detect low-amplitude variability on hour timescales.

Beyond its observational importance, INOV also provides a probe of the physical mechanisms driving rapid AGN variability. INOV studies of AGNs have shown that high-amplitude intranight variability is preferentially associated with radio-loud, jet-dominated systems. 
Blazars and related jet-hosting AGNs commonly exhibit INOV detection fractions of tens of percent and variability amplitudes exceeding 0.1 mag \citep[e.g.,][]{Goyal_etal_2013, Gupta_2018, Webb_etal_2021, Ojha_etal_2021}, likely driven by relativistic disturbances within jets \citep{Marscher_etal_1991, Calafut_Wiita_2015}. In contrast, radio-quiet AGNs generally exhibit weaker INOV signals in existing monitoring campaigns \citep{Kim_etal_2016, Negi2023, Yang_etal_2024}, with low-amplitude variations potentially arising from non-jet processes such as X-ray reprocessing or accretion-disk fluctuations \citep{Mangalam_Wiita_1993, Czerny_etal_2008, Dexter_Agol_2011, Cai_etal_2018, Sun_etal_2020}.

Given that IMBH candidates bridge the mass gap between stellar-mass BHs and SMBHs, they provide a unique regime for testing whether established accretion and variability scaling relations extend into the intermediate-mass range. Under disk-dominated variability scenarios, characteristic variability timescales are expected to scale with the central BH mass when compared at a fixed normalized disk radius. This expectation is supported by the observed dependence of the Damped Random Walk damping timescale ($\tau$) on BH mass across a broad range of accreting systems \citep{Burke2021, Zhou_etal_2024, Su_etal_2024}. Therefore, IMBH systems may exhibit variability on shorter characteristic timescales, motivating high-cadence monitoring campaigns.

However, detecting these compressed variability signals remains challenging. The observed optical variability of IMBH candidates can be substantially diluted by host-galaxy emission, reducing the apparent AGN variability amplitude.
Moreover, nearby IMBH candidates are often hosted by relatively prominent and spatially extended dwarf galaxies, where seeing-dependent PSF-fitting biases can introduce spurious variability signals comparable to the expected intrinsic fluctuations \citep{Krushinsky_etal_2025}. Therefore, reliable INOV measurements require robust photometric techniques, such as difference imaging, to mitigate seeing-dependent host-galaxy systematics and provide more reliable constraints on intrinsic variability.

Despite its importance for constraining accretion physics and optimizing future time-domain monitoring strategies, systematic statistical studies of INOV in the low-mass IMBH regime remain limited. A previous pioneering investigation reported an INOV detection fraction of $\sim$22\% for a sample of 12 low-mass AGNs \citep{Gopal-Krishna_etal_2023}. However, these targets were selected based on prior radio and X-ray detections, which may preferentially include systems with enhanced nuclear activity and higher AGN-to-host contrast. A larger optically selected sample without prior multi-wavelength selection is therefore required to establish a representative INOV constraint for the broader IMBH candidate population.

In this work, we investigate the occurrence and detectability of INOV in a large, optically selected sample of IMBH candidates using high-cadence observations from the Zwicky Transient Facility (ZTF), analyzed with difference-image photometry. To our knowledge, this represents a systematic investigation of INOV constraints in an optically selected IMBH candidate population by combining variability validation, long-term variability characterization, and Monte Carlo detectability simulations. Together, these analyses provide a comprehensive assessment of INOV and its detectability in this low-mass BH population. This paper is organized as follows. Section~\ref{sec:obs} describes the sample selection, ZTF data, and difference-image photometry procedure. Section~\ref{sec:verification} presents the INOV measurement and validation methodology. Section~\ref{sec:results} presents the observational results, including the INOV constraints and seeing-dependent variability properties. Section~\ref{sec:discussion} discusses the implications of our results for variability studies of low-mass BH systems, including the interpretation of the INOV non-detection, comparisons with previous studies, and future observational prospects. Section~\ref{sec:conclusion} summarizes our main conclusions. Throughout this paper, we adopt a flat $\Lambda$CDM cosmology with $H_{0}=70.0~\mathrm{km~s^{-1}~Mpc^{-1}}$ and $\Omega_{\rm m}=0.3$.

\section{Sample Selection and Difference-image Photometry}\label{sec:obs}
\subsection{Optical Light Curves from ZTF}\label{ztf}
The ZTF is a robotic time-domain survey conducted with a dedicated wide-field camera mounted on the Palomar 48-inch Schmidt telescope, providing large-scale optical monitoring with a field of view of $\sim$47~deg$^2$ and an 8~s readout time \citep{Bellm2019}. The public survey reaches median $5\sigma$ depths of 20.8 and 20.6~mag (AB; 30~s exposure) in the $g$ and $r$ bands, respectively \citep{Bellm2019, Masci_etal_2019}. The combination of long-baseline archival observations and dedicated high-cadence monitoring enables statistical investigations of optical variability over a wide range of timescales.

Because the $r$ band provides the densest temporal coverage and the largest number of intranight observations for our targets, we focus primarily on $r$-band light curves (LCs). These observations are collected from multiple ZTF public and partnership programs, including the Extragalactic High Cadence Survey, the High-Cadence Plane Survey, and the Twilight Survey \citep{Bellm2019, Masci_etal_2019}.

\subsection{Parent Sample}\label{sec:selection}
We compile a parent sample of 1,447 IMBH candidates from previous literature (Guo et al. 2026, in preparation), most of which were identified in SDSS-based surveys through broad H$\alpha$ emission and virial BH mass estimates. Among these candidates, 1,360 have publicly available SDSS spectra, which provide the basis for all spectroscopically derived parameters used in this work.

We re-analyze these spectra using a uniform spectral-decomposition procedure based on PyQSOFit \citep{Guo2018, Shen2019} to obtain homogeneous measurements of the key physical parameters used in this work, including the BH mass ($M_{\rm BH}$) and AGN flux fraction at 5100~\AA\ ($f_{\rm AGN}$). During the decomposition, a prior-based PCA host-galaxy modeling method \citep{Ren_etal_2024} is adopted to separate the AGN and host-galaxy components. The residual AGN continuum is modeled with a power-law component and iron emission, while emission lines are fitted with multiple Gaussian components. The velocity offsets and dispersions of narrow components are tied together, and the line centers are constrained according to laboratory wavelength separations.

For sources exhibiting broad H$\alpha$ emission, BH masses are estimated using single-epoch virial scaling relations based on the full width at half maximum (FWHM) and integrated luminosity ($L_{\mathrm{H}\alpha}$) of the broad-line component. The FWHM of broad H$\alpha$ is used as a tracer of the BLR gas velocity under the assumption of virialized motion dominated by the central BH, while $L_{\mathrm{H}\alpha}$ provides an estimate of the BLR radius through the empirical $R_{\mathrm{BLR}}$--$L_{\mathrm{H}\alpha}$ relation \citep{Greene2007}.

The AGN flux fraction, $f_{\mathrm{AGN}}$, is an important parameter for evaluating INOV detectability in low-luminosity IMBH candidates, where intrinsic AGN variability signals can be strongly diluted by dominant host-galaxy emission. In this work, it is defined as the ratio between the AGN continuum luminosity at 5100~\AA\ and the total continuum luminosity (AGN plus host galaxy) at the same wavelength, as obtained from spectral decomposition. We note that the derived $f_{\mathrm{AGN}}$ values are based on single-epoch SDSS spectra near rest-frame 5100~\AA\ and are used only as approximate proxies for the AGN contribution in the ZTF $r$ band. Differences in wavelength coverage, observing epoch, aperture, and seeing may introduce additional uncertainties into the inferred dilution correction.

\subsection{Target and Intranight Session Selection}\label{sec:session}
To construct a sample suitable for INOV measurements, we selected sources from the parent sample based on a series of photometric quality and observational criteria. First, we selected sources with mean $r$-band magnitudes brighter than $19~\mathrm{mag}$ over the ZTF survey duration to avoid regimes dominated by photon noise. Second, we retained only photometrically high-quality epochs with $\texttt{INFOBITS}=0$. Furthermore, because the ZTF pipeline processes observations from different fields, filters, and CCD quadrants independently, we selected only the primary LC corresponding to the observation ID with the largest number of available measurements to avoid artificial variability caused by combining heterogeneous LCs \citep{Negi2023}.

\begin{figure}[t!]
 \centering
 \includegraphics[width=\columnwidth]{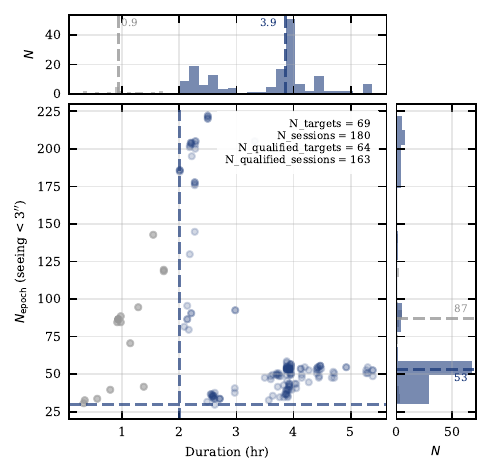}
 \caption{Duration and sampling statistics of the 180 candidate intranight sessions. Qualified sessions satisfy the criteria of $\geq30$ measurements with seeing $<3.0\arcsec$ and a continuous monitoring duration of $\geq2.0$~hr. The retained (163) and excluded (17) sessions are shown in blue and gray, respectively. \textbf{Main}: Number of qualified measurements as a function of session duration. Horizontal and vertical dashed lines indicate the adopted sampling and duration thresholds. \textbf{Top}: Distribution of session durations, with dashed lines marking the median values of the retained (3.9~hr) and excluded (0.9~hr) samples. \textbf{Right}: Distribution of the number of qualified measurements per session, with dashed lines indicating the medians of the retained (53) and excluded (87) samples.}
 \label{fig:duration_nfits}
\end{figure}

When convolved with the point spread function (PSF, which is dominated by atmospheric seeing for ground-based observations), the host galaxy of an AGN does not typically behave like a point source. Consequently, any intranight variation in the PSF can lead to fluctuations in the relative light contributions from the AGN and its host galaxy, particularly for low-$z$ candidates. To minimize such seeing-dependent effects, we imposed a maximum seeing threshold of $<3.0\arcsec$, corresponds to an approximately $1.5\times$ the typical ZTF median seeing \citep{Bellm2019}, while excluding the poor-seeing tail most affected by host-galaxy-induced systematics. After applying this criterion, we required each intranight session to contain at least 30 qualified measurements to ensure sufficient sampling for subsequent temporal binning and time-series analysis (See Section~\ref{sec:verification}). Restricting the sample to candidates with estimated BH masses below $10^6~M_{\odot}$, yielding an initial sample of 69 sources with 180 $r$-band intranight sessions, shown as black points in Figure~\ref{fig:duration_nfits}.

We further required each intranight session to have a continuous baseline duration of $\geq2$~hours. This criterion follows previous statistical INOV studies, which typically adopt minimum monitoring durations of $\sim2$--3 hours \citep{Kim_etal_2016, Gopal-Krishna_etal_2023, Negi2023, Ojha_etal_2024}, and ensures sufficient temporal coverage to distinguish low-amplitude variability trends from random photometric fluctuations. Since longer continuous monitoring windows improve sensitivity to short-timescale variability\citep[e.g.,][]{Gupta_etal_2005}, this criterion balances sample size and statistical reliability. The final science sample consists of 64 unique low-redshift IMBH candidates with 163 high-cadence $r$-band intranight sessions (maximum $z=0.188$), shown as blue points in Figure~\ref{fig:duration_nfits}.

\begin{figure}[t!]
 \centering
 \includegraphics[width=\columnwidth]{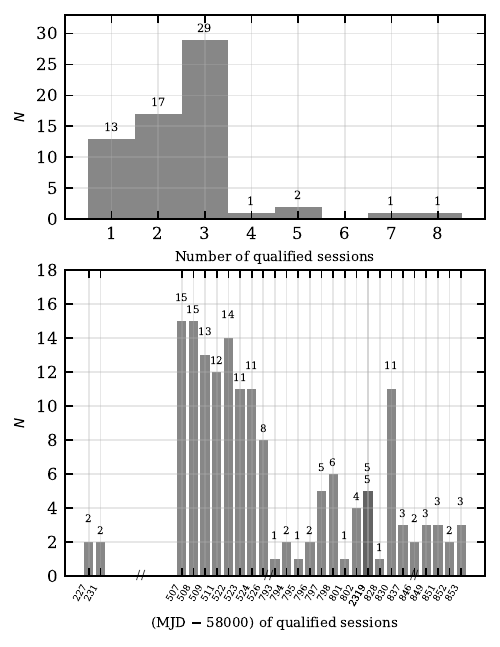}
 \caption{Distribution of the 163 qualified intranight sessions among the final sample of 64 IMBH candidates. \textbf{Top}: Number of intranight sessions contributed by each target, with the absolute number of targets annotated above each bin. \textbf{Bottom}: Temporal distribution of the available high-cadence sessions as a function of observing night ($\mathrm{MJD}-58000$).}
 \label{fig:session_stats}
\end{figure}

As shown in Figure~\ref{fig:duration_nfits}, the 163 qualified sessions contain a median of 53 measurements (with seeing $<3.0\arcsec$) and a median monitoring duration of 3.9~hours. The 17 excluded sessions were primarily removed because their durations were shorter than the 2-hour threshold, with a median duration of only 0.9~hours despite having a higher median sampling density of 87 measurements per session. Figure~\ref{fig:session_stats} summarizes the distribution of the 163 qualified intranight sessions among the final sample of 64 IMBH candidates. Most targets have one to three monitoring sessions (92.2\%), including 46.9\% with one or two sessions and 45.3\% with exactly three sessions, while only four targets (7.8\%) have four or more sessions.

\subsection{Forced Aperture Photometry on ZTF Difference Images}\label{sec:diff}
The ZTF archive provides $r$-band LCs generated by positional matching of sources detected across epochs in the PSF-fit photometry catalogs through the NASA/IPAC Infrared Science Archive (IRSA)\footnote{\url{https://irsa.ipac.caltech.edu/docs/program_interface/ztf_lightcurve_api.html}}. 
For host-dominated IMBH candidates, conventional PSF-fit photometry can be biased by seeing-dependent changes in the relative contributions of compact nuclear and extended host components.
To mitigate these systematics, we perform forced aperture photometry on ZTF difference images, which suppresses seeing-dependent host-galaxy contamination through reference-image subtraction.

To investigate both intranight and long-term variability, we retrieve the ZTF science, difference, and mask images from the IRSA image archive\footnote{\url{https://irsa.ipac.caltech.edu/ibe/}}. For the intranight LCs, we download image cutouts centered on each target for the monitoring sessions selected in Section~\ref{sec:session}, retaining only epochs with \texttt{infobits}=0 and seeing $<3\arcsec$. The downloaded difference images are those produced by the ZTF image-subtraction pipeline, which subtracts deep reference templates from individual science images after image registration, photometric scaling, and PSF matching. These reference templates are constructed from stacks of 15--40 high-quality historical CCD-quadrant images selected using astrometric, photometric, and image-quality criteria \citep{Masci_etal_2019}.

For the long-term LCs, following \citet{Arevalo_etal_2026}, we query the ZTF image metadata to retain only clean observations with \texttt{infobits}=0. We select the first epoch per night, requiring an MJD separation of $>0.5$~days for a given field, CCD quadrant, and filter combination, thereby constructing a homogeneous $\sim$7-year baseline while avoiding biases from densely sampled intranight observations.

Following the forced aperture photometry strategy of \citet{Zuo2024}, we use SExtractor in dual-image mode \citep{Bertin1996} instead of the {\tt Photutils} pipeline. The source positions are determined from the science images, and fluxes are subsequently extracted from the geometrically matched difference images. The initial comparison-star sample consists of five stars located on the same CCD chip as the target, selected to satisfy both spatial proximity ($<250\arcsec$) and photometric similarity ($\Delta m_r\leq1.0$ mag). If more than five candidates satisfy these criteria, the five stars with the closest $r$-band magnitudes to the target are adopted. Given the typical seeing of $<3\arcsec$ and the ZTF pixel scale of $1\arcsec$ pixel$^{-1}$, we adopt a circular aperture diameter of 6 pixels ($6\arcsec$), corresponding to approximately twice the maximum adopted seeing FWHM.

For both the target AGN and comparison stars, the total corrected instrumental fluxes ($f$) and uncertainties ($f_{\rm err}$) are derived by combining the difference-image and reference-image measurements:
\begin{subequations} 
\label{eq:cal_mprime}
\begin{align}
    f &= f_{\text{diff}} + f_{\text{ref}}, \\
    f_{\text{ref}} &= 10^{-0.4 \times (m_{\text{ref}} + m_{\text{zp,ref}} - m_{\text{zp,diff}})}, \\
    f_{\rm err} &= \sqrt{f_{\text{err,diff}}^2 + (1.0857 \times f_{\text{ref}} \times m_{\rm err,ref})^2}, \\
    m &= -2.5 \log_{10}(f) + m_{\text{zp,diff}}, \\
    m_{\rm err} &= 1.0857 \times \frac{f_{\rm err}}{f}.
\end{align}
\end{subequations}
Here, $m_{\rm ref}$ and $m_{\rm err,ref}$ represent the reference-image magnitude and its associated uncertainty measured within a $6\arcsec$ aperture, while $m_{\rm zp,ref}$ and $m_{\rm zp,diff}$ denote the photometric zero points of the reference and difference images, respectively. These comparison stars are used to empirically evaluate residual photometric systematics and validate the statistical significance of the target LCs (see details in Section~\ref{sec:verification}).

\section{INOV Measurement and Detection Procedure}\label{sec:verification}
\subsection{Light Curve Preparation and Quality Control}
Individual photometric measurements were first screened using a multi-step quality-control procedure, including iterative $5\sigma$ clipping to remove extreme outliers while minimizing potential impacts on variability measurements. The LCs were subsequently binned to reduce point-to-point photometric noise and improve the signal-to-noise ratio while maintaining sensitivity to variability on the monitored timescales. To ensure sufficient statistical sampling, each final binned LC was required to contain at least 10 valid measurements, consistent with the minimum sampling requirement adopted in Section~\ref{sec:session}.

To minimize potential dependence on a specific binning strategy, we adopted two independent binning approaches: (1) time-based binning with a characteristic bin width equal to three times the median separation between adjacent observations, and (2) grouping every three consecutive measurements. For both approaches, a new bin was initiated when the time separation between the current measurement and the latest point in the existing bin exceeded three times the median adjacent time separation. The consistency between the two binning methods provides a robustness check against binning-induced biases.

For each bin $i$ containing $N$ measurements, the binned magnitude ($m_i$) and its propagated uncertainty ($m_{i,\mathrm{err}}$) are calculated using an inverse-variance weighted mean:
\begin{subequations}
\label{eq:binning_weights}
\begin{align}
    m_i &= \frac{\sum_{k=1}^{N} w_k m_k}{\sum_{k=1}^{N} w_k},\\
    m_{i,\mathrm{err}} &= \left(\sum_{k=1}^{N} w_k\right)^{-1/2},
\end{align}
\end{subequations}
where $m_k$ and $m_{k,\mathrm{err}}$ represent the magnitude and uncertainty of the $k$-th measurement within the bin, respectively, and $w_k=1/m_{k,\mathrm{err}}^2$ is the inverse-variance statistical weight.

\subsection{Statistical Identification and Validation of INOV Candidates}
To preliminarily identify potential intranight variability in the LCs, we apply the commonly adopted $F$-test statistic for AGN INOV studies \citep{Diego2010}. For each binned LC, the $F$ value is calculated as
\begin{subequations} \label{eq:cal_Feta}
\begin{align}
    F &= \frac{\Delta_m}{\sigma^2_m},\\
    \Delta_m &= \frac{1}{N-1}\sum_{i=1}^{N}(m_i-\langle m\rangle)^2,\\
    \sigma^2_m &= \frac{1}{N}\sum_{i=1}^{N}m_{\mathrm{err},i}^{2},
\end{align}
\end{subequations}
where $m_i$ and $m_{\mathrm{err},i}$ are the magnitude and uncertainty of the $i$-th binned measurement in the intranight LC, respectively. $N$ is the number of binned measurements, $\Delta_m$ represents the observed variance of the LC, and $\sigma_m^2$ represents the expected variance from photometric uncertainties.

The calculated $F$ value is compared with a session-specific critical value ($F_c$) defined as the upper $99\%$ percentile ($\alpha=0.01$) of the nominal Snedecor $F$ distribution, corresponding to a nominal false-positive probability of $\sim1\%$ under the null hypothesis. The critical value is determined separately for each session according to the degrees of freedom of the binned LC ($N_{\rm bin}-1$), and sessions with $F>F_c$ are flagged as preliminary INOV candidates.

Five comparison stars located on the same CCD chip as the target are processed using the same difference-image photometry and variability analysis to validate candidate INOV detections. As these stars are assumed to be non-variable, their LCs provide a direct test of spurious variability. A target LC is considered a potential INOV detection only when it satisfies $F>F_c$ while all five comparison-star LCs satisfy $F<F_c$.

We further examine the relationship between the target LC and the seeing variation during each monitoring session. Sessions showing strong and statistically significant seeing correlations ($|r_{\rm Spearman}|>0.6$ and $p_{\rm Spearman}<0.05$) are excluded from the INOV candidate sample, as their apparent variability may be dominated by seeing-dependent photometric systematics associated with atmospheric changes and extended host-galaxy light, particularly for low-redshift candidates \citep{Cellone_etal_2007}.

Furthermore, for marginal or low-amplitude candidates where visual inspection is inconclusive, we adopt the temporal scatter measured from local comparison-star differential LCs as an empirical estimate of the photometric uncertainty. This accounts for additional measurement scatter not fully captured by the nominal SExtractor uncertainties, which may arise from reference-image stacking, image resampling, and PSF variations. Candidates that fail the $F$-test criterion after recalculating the $F$ statistic with this uncertainty estimate are not classified as robust INOV detections.

\subsection{Characterization of Candidate Variability Amplitudes}
For all monitoring sessions, we characterize the amplitude of the measured variability using two commonly adopted metrics: the peak-to-peak amplitude ($\phi$) in magnitude units \citep{Heidt_Wagner1998} and the fractional variability amplitude ($V$), which accounts for measurement uncertainties and provides a noise-corrected estimate of the variability strength \citep{Vaughan2003}. These metrics are calculated for all the sessions to characterize the variability amplitude distribution, but are not adopted as independent detection criteria. The definitions of $\phi$ and $V$ are provided in Appendix~\ref{app:var_amp}.

\section{Results}\label{sec:results}
\subsection{Statistical Constraints on INOV}\label{sec:INOV}
\begin{figure*}[t!]
 \centering
 \includegraphics[width=\textwidth]{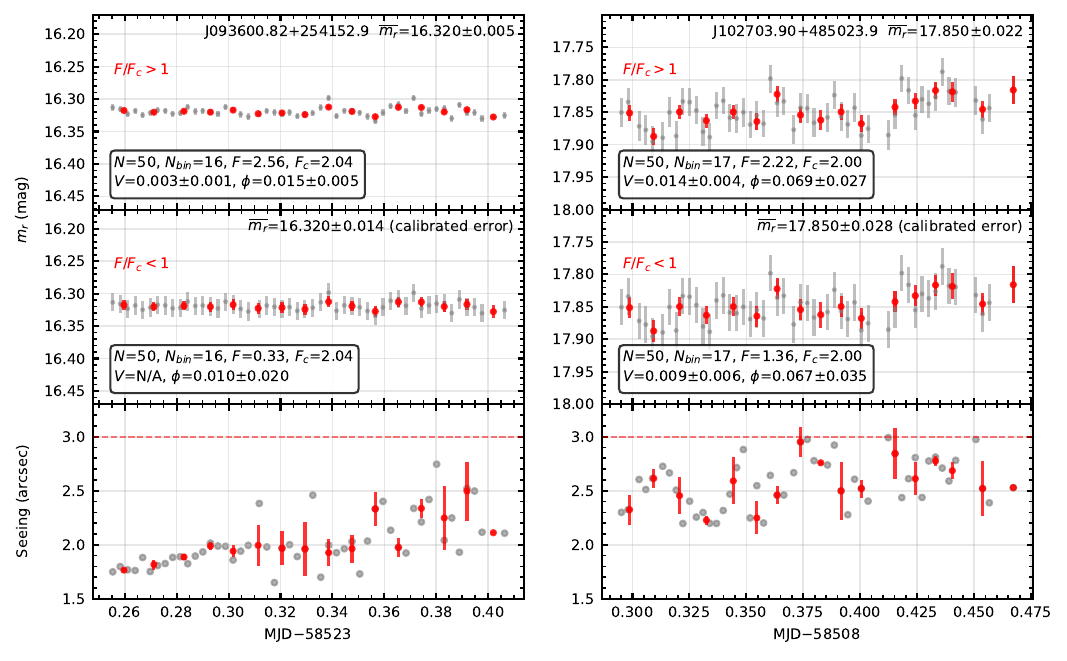}
 \caption{Photometric error calibration for the two sessions initially flagged as showing spurious INOV. \textbf{Top}: Binned LCs adopting nominal SExtractor uncertainties ($\sim$0.005--0.022\ mag), which falsely yield $F/F_c > 1$. \textbf{Middle}: The same LCs with uncertainties estimated from the median temporal scatter of local comparison-star differential LCs ($\sim$0.014--0.028~mag), resulting in $F/F_c$ values below the variability threshold. \textbf{Bottom}: The atmospheric seeing during each session.}
 \label{fig:error_calibration}
\end{figure*}

The statistical variability analysis applied to our SExtractor dual-image difference LCs initially identified four and eight sessions exhibiting apparent variability under binning methods 1 and 2, respectively. Two sessions were common to both approaches: J$093600+254153$ (MJD 58523) and J$102703+485024$ (MJD 58508). However, the variability amplitudes are small, with peak-to-peak amplitudes of only $\phi\sim0.02$--$0.07$ mag. 
As shown in Figure~\ref{fig:error_calibration}, although these sessions satisfy the variability criterion ($F/F_c>1$) when using the nominal SExtractor uncertainties, they no longer pass the $F$-test when the target uncertainties are recalculated using the median temporal scatter of local comparison-star differential LCs within the same monitoring session. This empirical uncertainty estimate reflects the typical photometric scatter of non-variable sources under identical observing conditions. This indicates that these tentative detections are more likely associated with underestimated nominal uncertainties than with genuine intrinsic INOV.

\begin{figure}[t!]
 \centering
 \includegraphics[width=\columnwidth]{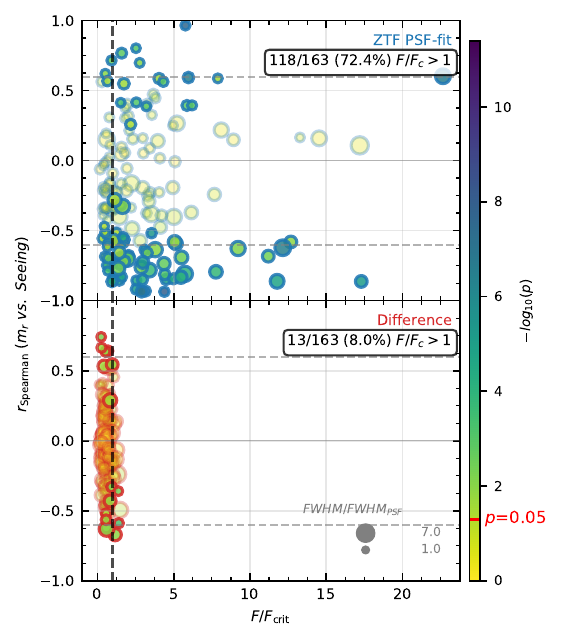}
 \caption{Seeing dependence of the apparent variability significance for the 163 intranight sessions. The horizontal axis shows $F/F_c$, and the vertical axis shows the magnitude-seeing Spearman correlation coefficient ($r_{\rm Spearman}$). Marker sizes represent the source extension ratio (${\rm FWHM/FWHM_{\rm PSF}}$), measured relative to the median stellar PSF on the same CCD chip, while colors indicate the significance of the seeing correlation through $-\log_{10}(p_{\rm Spearman})$. The vertical and horizontal dashed lines mark the variability threshold ($F/F_c=1$) and seeing-correlation criterion ($|r_{\rm Spearman}|=0.6$), respectively. \textbf{Top}: ZTF PSF-fit LCs. \textbf{Bottom}: SExtractor dual-image difference LCs.}
\label{fig:correlation_vs_seeing}
\end{figure}

Figure~\ref{fig:correlation_vs_seeing} presents the ensemble behavior of all 163 sessions under binning method 1 by showing the magnitude-seeing correlation coefficient ($r_{\mathrm{Spearman}}$) as a function of $F/F_c$. The upper panel shows the ZTF PSF-fit LCs, while the lower panel presents our SExtractor dual-image difference LCs. The data points are size-coded according to the source extension ratio (${\rm FWHM/FWHM_{PSF}}$), where the target FWHM is normalized by the median stellar FWHM measured from stars satisfying \texttt{CLASS\_STAR}$>0.95$, \texttt{FLAGS}$=0$, and \texttt{MAGERR\_BEST}$<0.03$ mag on the same CCD chip. The extension ratios range from 1.18 to 12.65, with 66\% of sessions exceeding 1.5, demonstrating that most targets exhibit measurable extension relative to the stellar PSF.

The two panels show a clear difference in their apparent variability distributions. For the ZTF PSF-fit LCs, 72.4\% (118/163) of the sessions satisfy $F/F_c>1$. In contrast, the dual-image difference LCs exhibit a strong shift toward lower $F/F_c$ values, with an ensemble median of $F/F_c=0.57\pm0.27$, where the uncertainty represents the $1\sigma$ dispersion around the median. This reduction decreases the apparent variability fraction to 8.0\% (13/163).

These apparent candidates were then subjected to the full INOV validation pipeline. Removing sessions with significant seeing correlations reduces the number of surviving candidates to 76 and 12 for the PSF-fit and difference-image LCs, respectively. Requiring consistency between the two binning methods further reduces these numbers to 61 and 10. After applying the five-comparison-star validation and visual inspection, 55 PSF-fit and 2 difference-image candidates remain for further inspection, with the latter two subsequently rejected after comparison with empirical uncertainty estimates. Thus, the excess apparent INOV detected by conventional PSF-fit photometry is primarily caused by seeing-dependent host-galaxy systematics, while difference-image photometry mitigates these effects and yields no robust INOV detections.

\subsection{Seeing-dependent Host-galaxy Effects}
\subsubsection{Representative Examples}\label{sec:seeing_examples}
To illustrate the impact of atmospheric seeing on ZTF PSF-fit LCs, we examine two representative sessions showing opposite magnitude-seeing correlation trends. These sessions are selected from sources flagged as variable by the ZTF PSF-fit LCs but classified as non-variable by our difference-image analysis. We select two representative cases showing the strongest positive (Example~A) and negative (Example~B) correlations between PSF-fit magnitude and seeing, with $r_{\rm Spearman}=+0.59$ ($p=3.8\times10^{-7}$) and $r_{\rm Spearman}=-0.58$ ($p=0.011$), respectively.

\begin{figure}[t!]
 \centering
 \includegraphics[width=\columnwidth]{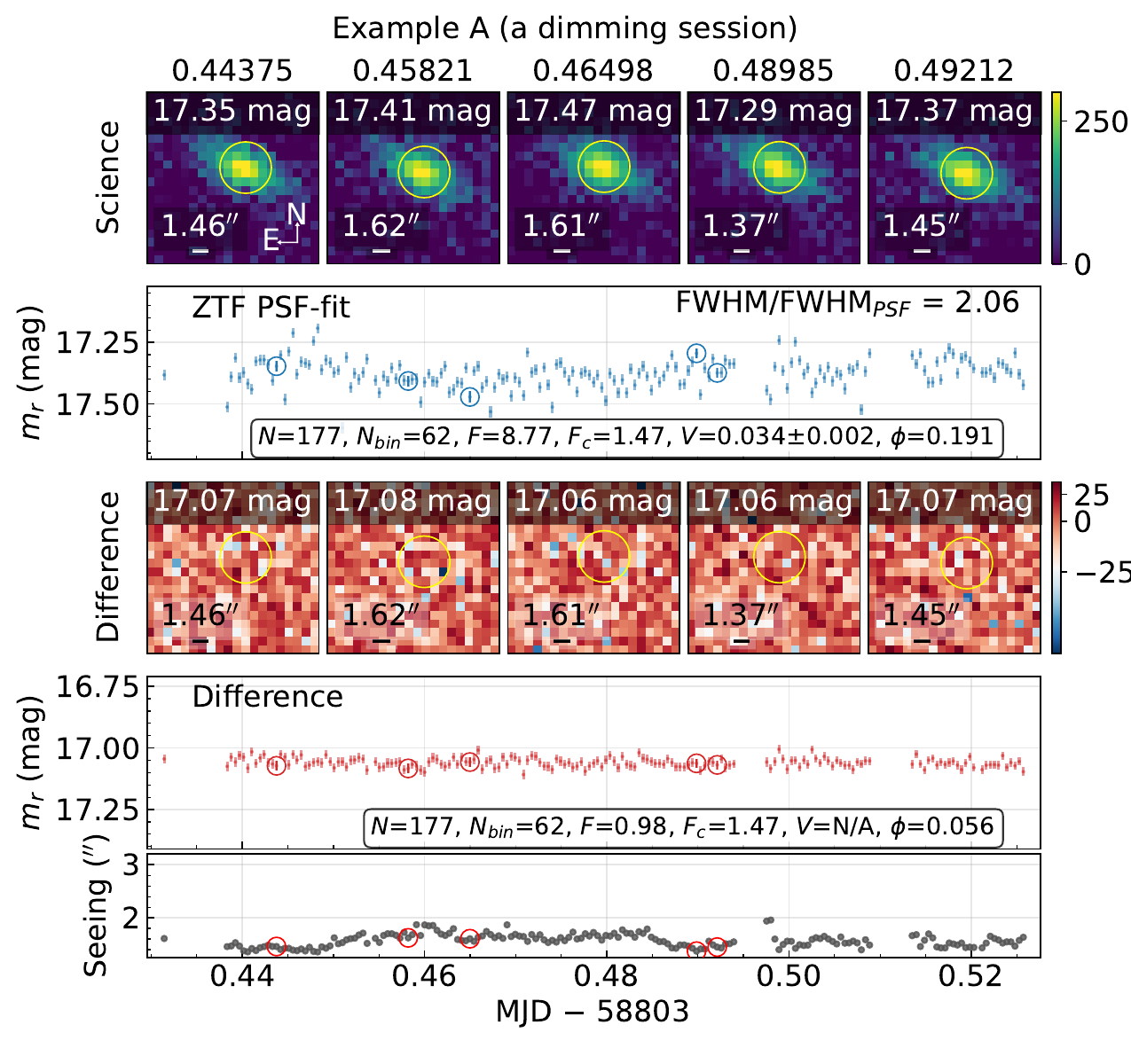}
 \caption{Diagnostic visualization of Example~A (a dimming session) for $\mathrm{J0914+1156}$ on MJD=58803. Columns correspond to five representative epochs selected from the 5th, 28th, 50th, 72nd, and 95th percentiles of the nightly ZTF PSF-fit magnitude distribution. \textbf{Row 1}: Background-subtracted ZTF science-image cutouts ($20\arcsec\times20\arcsec$; \texttt{asinh} stretch), with the seeing scale bar and adopted $6\arcsec$ aperture diameter indicated. \textbf{Row 2}: ZTF PSF-fit LC with the selected epochs marked. \textbf{Row 3}: Reference-subtracted difference-image cutouts centered on zero with a per-image $\pm3\sigma$ \texttt{asinh} stretch. \textbf{Row 4}: SExtractor dual-image difference LC. \textbf{Row 5}: Time-resolved seeing variation during the session.}
 \label{fig:details_1}
\end{figure}

\begin{figure}[t!]
 \centering
 \includegraphics[width=\columnwidth]{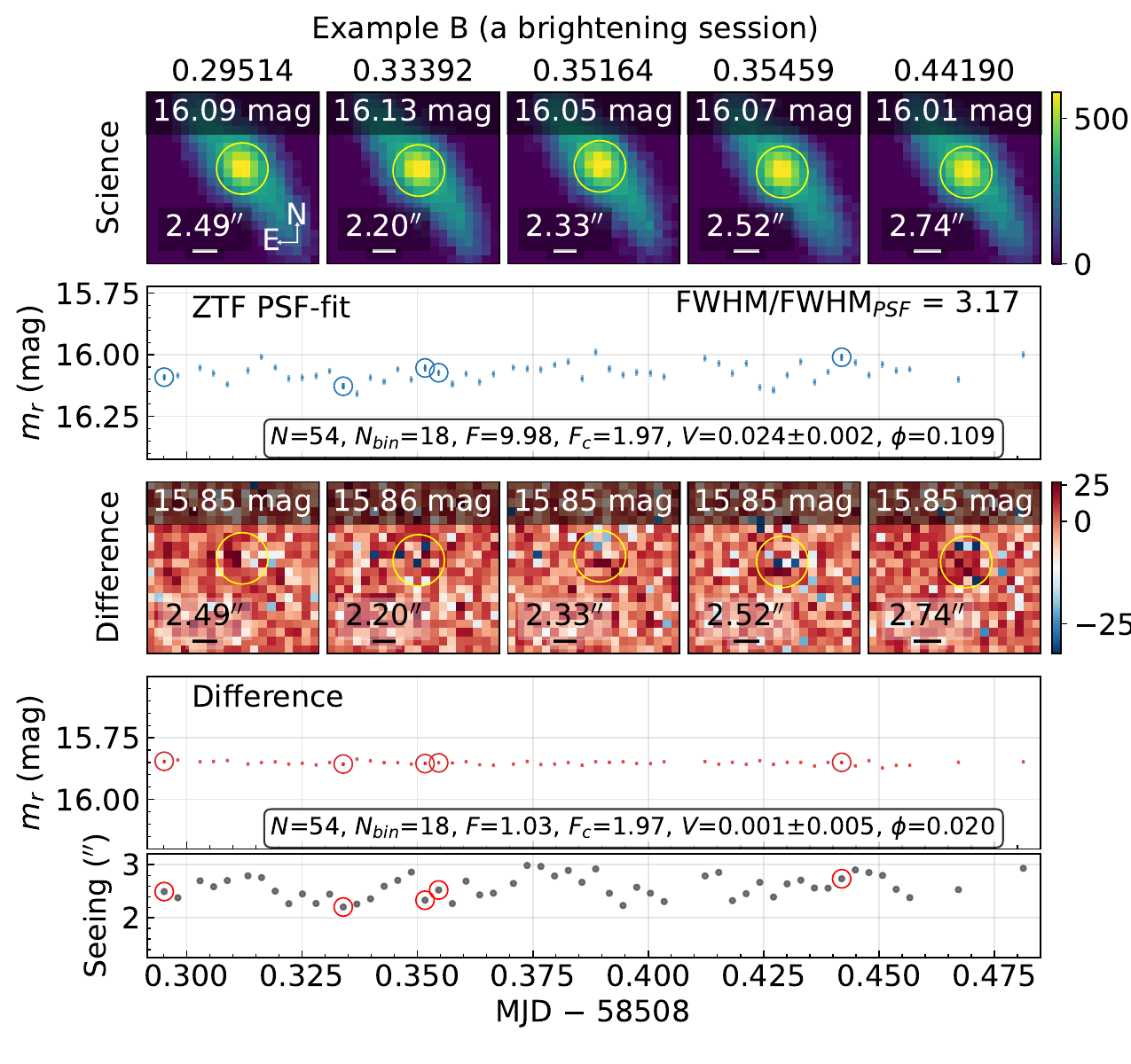}
 \caption{Same as Figure~\ref{fig:details_1}, but for Example~B (a brightening session) of $\mathrm{J1048+5002}$ on MJD=58508. The session shows an opposite magnitude-seeing trend in the ZTF PSF-fit LC compared with Example~A, while the SExtractor dual-image difference LC remains stable.}
 \label{fig:details_2}
\end{figure}

For each session, five representative epochs were selected at the 5th, 28th, 50th, 72nd, and 95th percentiles of the clipped nightly PSF-fit LC flux distribution, sampling the full range of observed flux states while avoiding the most extreme values. The epochs with the minimum and maximum seeing among these five measurements are adopted as the best- and worst-seeing cases, respectively. Figures~\ref{fig:details_1} and \ref{fig:details_2} show the corresponding science and difference-image cutouts together with the photometric diagnostics.

To investigate the origin of the seeing-dependent PSF-fitting bias, we further examine the spatial light profiles of the two representative sources (see Appendix~\ref{app:SB_profiles}).
We characterize the central light concentration using two structural metrics: the concentration index, defined as the enclosed flux ratio within $1\arcsec$ and $3\arcsec$ apertures, and the peak-to-baseline contrast, defined as the ratio between the surface brightness at $0$--$0.5\arcsec$ and $5$--$5.5\arcsec$ radii. These quantities characterize the degree of central light concentration and the relative contrast between compact nuclear and extended host-galaxy components.

Both examples exhibit a trend toward lower central concentration under poorer seeing conditions ($0.26\rightarrow0.25$ for Example~A and $0.24\rightarrow0.21$ for Example~B), consistent with the expected broadening of the observed central light profile. This behavior is also visible in the science-image cutouts (top row of both figures), where the targets appear increasingly diffuse under poorer seeing conditions.

Example~A exhibits a dimming of $0.12$ mag as the seeing increases by $\sim0.25\arcsec$. This source shows a high peak-to-baseline contrast at the best-seeing epoch ($35.6$), which decreases to $29.2$ under poorer seeing, indicating that the centrally concentrated component becomes diluted as the PSF broadens. In contrast, Example~B brightens by $0.12$ mag as the seeing increases by $\sim0.54\arcsec$. 
With a lower initial contrast ($14.9$), this source has a larger relative contribution from extended host emission and shows the opposite photometric response to seeing variations. Together, these two examples illustrate that seeing-induced photometric biases can produce either apparent dimming or brightening, depending on the relative dominance of compact nuclear and extended host emission.

As shown in Figures~\ref{fig:details_1} and \ref{fig:details_2}, despite the strong seeing-dependent trends in the ZTF PSF-fit LCs, the SExtractor dual-image difference LCs remain nearly flat, with $F/F_c\sim0.67$ for Example~A (J0914+1156) and $F/F_c\sim0.52$ for Example~B (J1048+5002). Although weak residual magnitude-seeing correlations may remain due to imperfect PSF matching, the much lower variability significance in the difference-image LCs indicates that difference-image aperture photometry strongly reduces seeing-dependent host-galaxy systematics.

\subsubsection{Statistical properties}\label{sec:seeing}
To obtain a statistical view of seeing-induced magnitude variations across the full sample, we extend the case-study analysis (Section~\ref{sec:seeing_examples}) to all 163 qualified monitoring sessions. For each session, the best- and worst-seeing epochs are selected from the five percentile-based epochs defined in Section~\ref{sec:seeing_examples}. We quantify the corresponding PSF-fit magnitude response as $\Delta m=m_{\rm worst}-m_{\rm best}$ and compare it with the mean photometric uncertainty of the two measurements.

\begin{figure}[t!]
 \centering
 \includegraphics[width=\columnwidth]{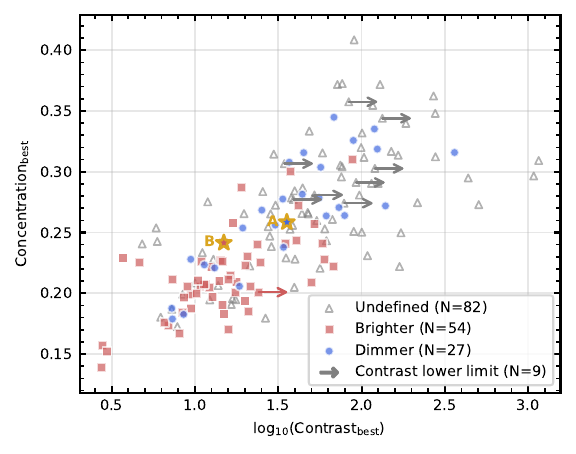}
 \caption{Concentration index versus peak-to-baseline contrast measured at the best-seeing epoch for all 163 ZTF intranight sessions. Sessions are classified according to the ZTF PSF-fit magnitude response between the best- and worst-seeing epochs: dimming sessions (blue circles, $N=27$), brightening sessions (red squares, $N=54$), and sessions without significant seeing-dependent magnitude variations ($|\Delta m|<3\bar{m}_{\rm err}$; gray triangles, $N=82$). Gold stars indicate the two representative case studies (Examples~A and B). Dimming sessions preferentially occupy the high-concentration/high-contrast region, whereas brightening sessions are concentrated toward lower values of both parameters. Rightward arrows indicate sessions with sky-noise-limited host baseline measurements, for which the peak-to-baseline contrast represents a lower limit.}
 \label{fig:mechanism}
\end{figure}

Using a $3\sigma$ threshold, we classify each session into three empirical categories: (1) Dimming ($\Delta m>+3\bar{m}_{\rm err}$), where the source becomes fainter under poorer seeing; (2) Brightening ($\Delta m<-3\bar{m}_{\rm err}$), where the source becomes brighter under poorer seeing; and (3) Stable/Undefined ($|\Delta m|\leq3\bar{m}_{\rm err}$), where no significant seeing-dependent response is detected. This classification yields 27 dimming, 54 brightening, and 82 stable/undefined sessions. For targets where the outer host-galaxy baseline annulus ($5\arcsec$--$5.5\arcsec$) is limited by sky noise, the peak-to-baseline contrast is conservatively floored using the $1\sigma$ sky-noise level, producing the lower limits indicated by arrows in Figure~\ref{fig:mechanism}.

The structural measurements reveal a systematic decrease in central concentration under degraded seeing conditions. A decrease in concentration index is observed in 93\% (25/27) of dimming sessions and 94\% (51/54) of brightening sessions, with both populations showing a median decrease of approximately 10\%.
This confirms that worsening seeing consistently broadens the observed light profiles. As shown in Figure~\ref{fig:mechanism}, the dimming and brightening populations occupy statistically distinct regions of the contrast--concentration plane. The dimming group exhibits higher peak-to-baseline contrast and concentration values (median values of $36.8\pm70.7$ and $0.27\pm0.05$), whereas the brightening group shows lower values ($14.8\pm17.9$ and $0.21\pm0.03$). A two-sample KS test on the contrast distributions gives $D=0.52$ ($p\sim10^{-4}$), indicating a significant difference between the two populations. Similarly, brightening sessions exhibit larger source extension ratios (${\rm FWHM/FWHM}_{\rm PSF}=2.52\pm2.13$) than dimming sessions ($1.73\pm0.86$), with the corresponding KS test rejecting identical distributions ($D=0.43$, $p\sim2\times10^{-3}$).

Nevertheless, morphology alone cannot uniquely predict the direction of the seeing-induced magnitude response. Using the midpoint of the median contrasts ($\sim26$) as a simple separator, 19 of the 81 classified sessions ($\sim23\%$) deviate from the expected trend, including low-contrast dimming cases and high-contrast brightening cases. A logistic regression model based only on contrast correctly classifies 59/81 sessions (73\%), only modestly exceeding the 67\% majority-class baseline obtained by assigning all sessions to the brightening category. The concentration index alone provides no additional predictive power. Therefore, although central concentration and host dominance influence the magnitude response, the observed behavior likely depends on additional factors, including the amplitude of seeing variations, PSF-fitting details, and the detailed morphology of individual host galaxies.

Our sample is strongly host-dominated, with $f_{\rm AGN}$ derived from spectral decomposition (Section~\ref{sec:selection}) ranging from $\sim0$ to $0.25$ and a median value of $\sim0.06$. In such systems, even modest atmospheric seeing variations can produce PSF-fit magnitude fluctuations comparable to or exceeding typical INOV amplitudes through changes in the recovered AGN-host light distribution. These results show that difference-image photometry provides a more robust assessment of short-timescale variability by reducing seeing-dependent host-galaxy systematics in low-redshift, host-dominated IMBH candidates.

\section{Discussion}\label{sec:discussion}
\subsection{Interpretation of the INOV Detection Fraction}\label{sec:detection_fraction}
Among the 64 sources with suitable ZTF monitoring data, none exhibits robust INOV after applying our full validation procedure. To quantify the incidence of detectable INOV signals in our monitoring sample, we adopt the session-level detection fraction commonly used in previous INOV studies \citep{Romero_etal_1999}:
\begin{equation}
    f_{\rm INOV} =
    100\% \times
    \frac{\sum_{i=1}^{n} N_i(1/\Delta T_i)}
    {\sum_{i=1}^{n}(1/\Delta T_i)},
\end{equation}
where $\Delta T_i=\Delta T_{i,\rm obs}/(1+z)$ is the rest-frame duration of the $i$-th monitoring session, and $N_i$ is a binary indicator that equals 1 for a confirmed INOV detection and 0 otherwise.
In our sample, all 163 monitoring sessions have $N_i=0$, resulting in an empirical session-level detection fraction of zero. With 163 monitoring sessions, the minimum non-zero session-level detection fraction measurable with our dataset is $1/163\approx0.6\%$. This value should not be interpreted as a formal statistical upper limit, but rather as the sensitivity scale of the session-level detection fraction.

\subsection{Linking long-term variability to INOV detectability} \label{sec:longterm}
To place the INOV constraint in the context of longer-timescale stochastic variability, we extend our analysis to the multi-year variability properties of the sample. Using the same difference-image photometry framework, we construct long-term LCs and characterize the variability amplitudes and timescales through ensemble structure function (SF) analysis, providing an empirical reference for interpreting the detectability of short-timescale variability.

\subsubsection{Light Curve Construction and Variable Sample Selection}
For each of the 64 selected targets, we constructed long-term $r$-band LCs spanning the $\sim$7-year ZTF archival baseline using the SExtractor dual-image forced photometry framework described in Section~\ref{sec:diff}. To ensure consistency with the intranight analysis, the long-term LCs were processed using the same photometric quality control procedure and iterative $5\sigma$ clipping to remove extreme outliers.

Because the ZTF temporal sampling is non-uniform, we applied an adaptive binning procedure to obtain robust long-term variability measurements. The binning scale was determined individually for each target, typically ranging from 5 to 10 days with a median value of 7 days (Appendix~\ref{app:longterm_binning}).

\begin{figure*}[t!]
\centering
\includegraphics[width=\textwidth]{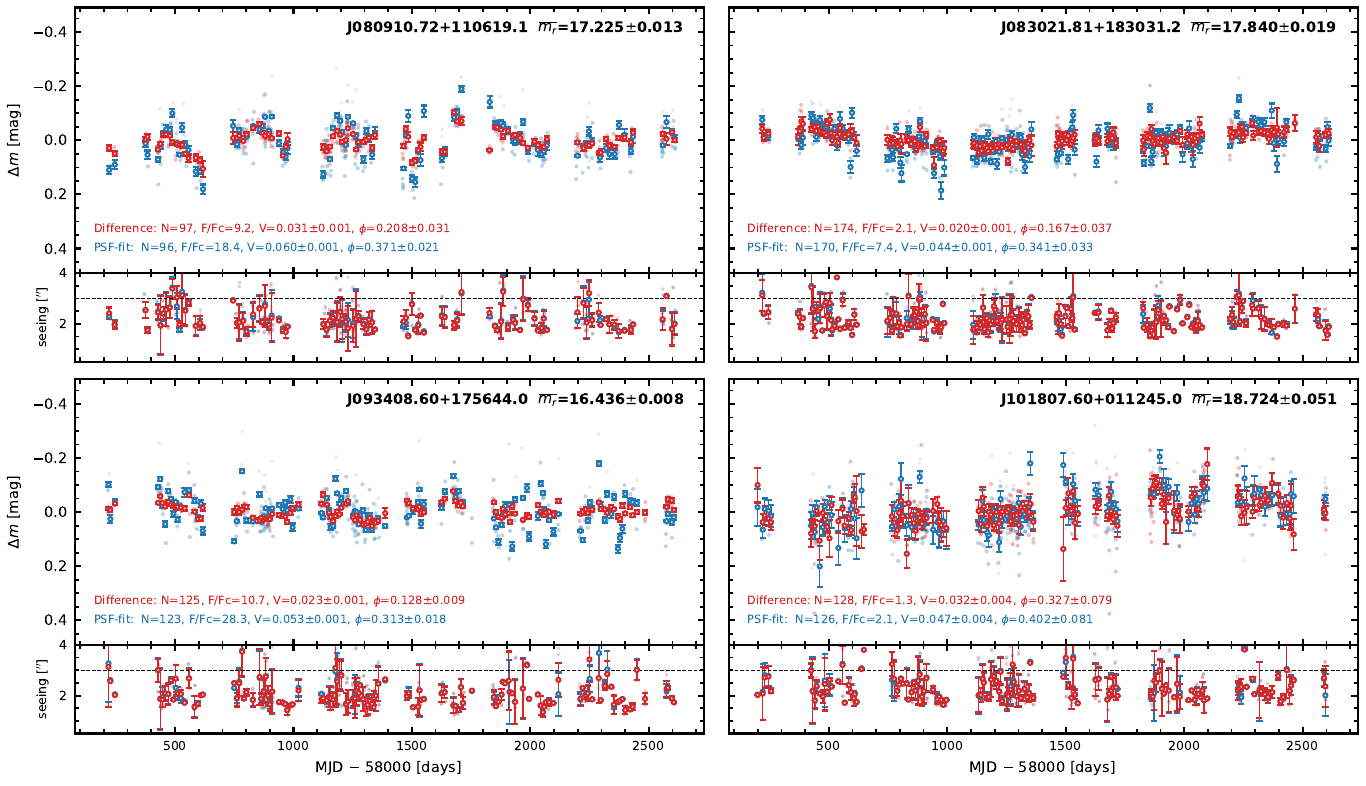}
\caption{$r$-band LCs of the four high-confidence long-term variable IMBH candidates identified from the ZTF difference-image analysis. Each panel shows the archival PSF-fit LC (blue points) together with the difference-image LC constructed in this work and its adaptively binned measurements (open squares). The lower sub-panels display the corresponding nightly seeing measurements and their binned values. Panel headers indicate the source designation and the median archival $r$-band magnitude.}
\label{fig:variable_lc_showcase}
\end{figure*}

We identify long-term variable candidates by first applying statistical selection based on the $F$-test significance, fractional variability amplitude $V$, and peak-to-peak variability amplitude $\phi$, yielding 23 candidates among the 64 targets. For the subsequent ensemble SF analysis, we further refine this sample through visual inspection, retaining only sources exhibiting coherent long-term variability over the full ZTF baseline (see Appendix~\ref{app:longterm_selection}). Four targets---J080910.72+110619.1, J083021.81+183031.2, J093408.60+175644.0, and J101807.60+011245.0---are adopted as the high-confidence long-term variable subsample, and their difference-image LCs are shown in Figure~\ref{fig:variable_lc_showcase}. The remaining 60 targets are used as the variability-undetected reference sample for the ensemble SF analysis, providing an estimate of the observed photometric noise floor.

\subsubsection{Ensemble Structure Function Construction}
Taking advantage of the combined intranight and long-term ZTF coverage of our sample, we construct ensemble SFs spanning intranight to multi-year timescales ($\Delta t\sim0.003$--$1600$ days). The intranight regime is obtained from within-session epoch pairs, while the long-term regime is derived from the binned difference-image LCs. Together, these measurements provide variability measurements over a continuous range of timescales.

For each subsample, the ensemble SF is calculated by combining all unique epoch pairs within individual sources:
\begin{equation}
{\rm SF}(\Delta t)=
\sqrt{\left\langle \Delta m^2 \right\rangle-
\left\langle \sigma_{\rm comb}^2 \right\rangle},
\end{equation}
following the standard noise-corrected definition of the SF \citep[e.g.,][]{Press_etal_1992,Kozlowski2016}. Here, $\Delta m=m_i-m_j$ is the magnitude difference for an epoch pair, $\sigma_{\rm comb}^2=\sigma_i^2+\sigma_j^2$ is the corresponding measurement-noise variance assuming independent measurement errors, and $\sigma_i$ and $\sigma_j$ are the estimated photometric uncertainties of the two measurements. The angle brackets denote the mean over all epoch pairs within a given lag bin. The corresponding noise-inclusive root-mean-square (RMS) dispersion is simply $\sqrt{\langle\Delta m^2\rangle}$. The SF uncertainties are estimated using source-based bootstrap resampling (see Appendix~\ref{app:sf_construction}).

\subsubsection{Structure Function Modeling}
\begin{figure}[t!]
\centering
\includegraphics[width=\columnwidth]{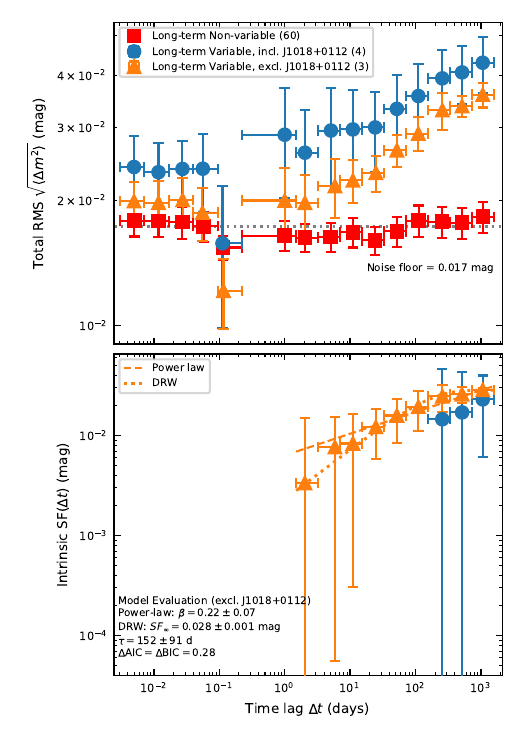}
\caption{Ensemble SF analysis. \textbf{Top}: Noise-inclusive RMS dispersion, $\sqrt{\langle\Delta m^2\rangle}$, for the variability-undetected reference subsample (red squares; 60 sources), the long-term variable subsample including J1018+0112 (blue circles; 4 sources), and the primary three-source modeling subsample excluding J1018+0112 (orange triangles; 3 sources). The horizontal dotted line indicates the observed photometric noise floor ($\sim$0.017~mag) derived from the variability-undetected reference subsample. \textbf{Bottom}: Noise-subtracted ensemble SF for the two definitions of the long-term variable subsample. The dotted and dashed curves show the DRW ($\tau=152\pm91$~days, ${\rm SF}_{\infty}=0.0285$~mag) and power-law ($\beta=0.22\pm0.07$) models fitted to the three-source modeling subsample over $\Delta t=1.5$--$1600$~days. Horizontal bars indicate the adopted time-lag bin widths.}
\label{fig:ensemble_SF}
\end{figure}

As shown in Figure~\ref{fig:ensemble_SF} (upper panel), the variability-undetected reference subsample establishes a stable observed photometric noise floor of $\sim0.017$ mag. The selected variable subsample exhibits systematically larger noise-inclusive RMS dispersions, reaching $\sim0.02$--$0.024$ mag at the shortest sampled timescales and increasing to $\sim0.036$--$0.043$ mag at $\Delta t\sim800$--$1000$ days.

The lower panel presents the ensemble SF for the selected variable subsample. Lag bins with $\left\langle\Delta m^2\right\rangle \leq \left\langle\sigma_{\rm comb}^2\right\rangle$ are omitted from the lower panel but remain included in the noise-inclusive RMS dispersion shown in the upper panel.

Target J1018+0112, the faintest of the four sources, has substantially larger difference-image photometric uncertainties, resulting in fewer lag bins with measurable ensemble SF after noise subtraction. We therefore adopt the three-source subsample excluding J1018+0112 for the formal SF modeling, while presenting both the three- and four-source results for comparison. For the adopted three-source subsample, the ensemble SF increases from $\sim0.003$ mag at $\Delta t\sim2$ days and approaches an asymptotic amplitude of ${\rm SF}_{\infty}\sim0.0285$ mag by $\Delta t\sim500$ days.

To quantify the characteristic variability timescale, we fit two commonly adopted models to the ensemble-SF measurements over the $\Delta t=1.5$--$1600$ day baseline: a power-law model \citep[e.g.,][]{Schmidt_etal_2010},
\begin{equation}
    {\rm SF}(\Delta t) = C\,\Delta t^{\beta},
    \label{eq:sf_pl}
\end{equation}
and a damped random walk (DRW) model \citep{Kelly_etal_2009},
\begin{equation}
    {\rm SF}(\Delta t) = {\rm SF}_\infty
    \sqrt{1-e^{-\Delta t/\tau}},
    \label{eq:sf_drw}
\end{equation}
where $C$ and $\beta$ are the normalization and slope of the power-law model, respectively, while ${\rm SF}_\infty$ and $\tau$ are the asymptotic variability amplitude and characteristic damping timescale of the DRW model. The model parameters are determined by minimizing the weighted $\chi^2$ using the bootstrap-derived SF uncertainties.

The two models are compared using the Akaike and Bayesian Information Criteria (AIC and BIC). The DRW model yields an asymptotic amplitude of ${\rm SF}_{\infty}=0.0285\pm0.0011$ mag and a turnover timescale of $\tau=152\pm91$ days ($\chi^2=0.39$). However, its statistical performance is comparable to that of the power-law model ($\beta=0.22\pm0.07$, $\chi^2=0.67$), with $\Delta{\rm AIC}=\Delta{\rm BIC}=0.3$.

The small differences in AIC and BIC indicate that the current data do not provide sufficient evidence to distinguish between the DRW and power-law descriptions. The flattening of the ensemble SF at long lags is consistent with a DRW-like turnover, while the shallow power-law slope ($\beta\sim0.22$, corresponding to a PSD slope of $\alpha\sim1.4$) is consistent with red-noise variability with enhanced power toward longer timescales.

The best-fit DRW turnover timescale ($152\pm91$ days), although poorly constrained, is considerably longer than the $\sim8$--19 day range expected for BH masses of $10^5$--$10^6,M_\odot$ from the empirical BH mass--timescale relation of \citet{Burke2021}. Since this estimate is derived from the small long-term variable subsample used for the ensemble SF analysis, it should not be regarded as characteristic of the overall IMBH population. Whether this apparent discrepancy reflects sample-selection effects or a departure from the canonical DRW scaling relation remains unclear.

Finally, the ensemble SF is not measurable at intranight timescales ($\Delta t\lesssim0.2$ days) with the current photometric precision, but becomes measurable at multi-day lags. This suggests that the transition to detectable intrinsic variability occurs over approximately $\Delta t\sim0.2$--$2$ days, providing an empirical basis for interpreting the absence of detectable INOV in our sample.

\subsection{Simulation}\label{sec:simulation}
To quantify the detectability of INOV in IMBH candidates, we perform Monte Carlo simulations to estimate the probability of recovering short-timescale variability under realistic ZTF observing conditions. Although difference-image photometry mitigates seeing-dependent host-galaxy systematics, the intrinsic AGN variability remains physically diluted by host-galaxy light in the observed flux. We therefore evaluate INOV recovery by incorporating the expected variability amplitudes, AGN fractions, ZTF cadence, and photometric uncertainties.

\subsubsection{Simulation Method}
We model the intrinsic AGN variability as a DRW, an Ornstein--Uhlenbeck stochastic process characterized by a damping timescale $\tau$ and a driving amplitude $\sigma_{\rm DRW}$. Following the normalization convention of \citet{Kelly_etal_2009}, the asymptotic SF amplitude is related to the DRW parameters as
\[
{\rm SF}_{\infty}=\sigma_{\rm DRW}\sqrt{\tau}.
\]
We scale $\tau$ with BH mass using the empirical relation from \citet{Burke2021},
\[
\tau = 107 \times (M_{\rm BH}/10^8M_\odot)^{0.38}\ {\rm days},
\]
which gives $\tau=7.8$, $12.6$, and $18.6$~days for $\log(M_{\rm BH}/M_\odot)=5.0$, $5.5$, and $6.0$, respectively.

For each target, extended DRW realizations are first generated using the recursive realization algorithm described by \citet{Kelly_etal_2009}. To minimize dependence on initial conditions and edge effects \citep{Kelly_etal_2009, MacLeod_etal_2010}, each realization is generated on a 1-minute time grid as a master LC spanning at least $10\tau$ ($78$--$186$~days). The observed monitoring segments are then randomly extracted from these realizations and sampled according to the adopted survey cadence.

\begin{figure}[t!]
 \centering
 \includegraphics[width=\columnwidth]{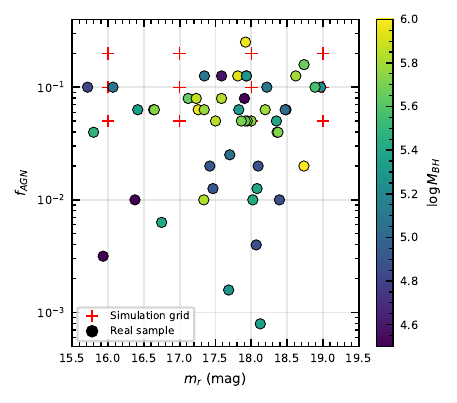}
 \caption{Distribution of the 50 targets with reliable $f_\mathrm{AGN}$ measurements in the $m_r$--$f_\mathrm{AGN}$ plane, color-coded by $\log M_\mathrm{BH}$. The sample is concentrated toward faint magnitudes and low AGN contributions, with 82\% of targets having $m_r>17$ mag and 74\% having $f_\mathrm{AGN}<0.10$. The remaining 14 targets lacking reliable $f_\mathrm{AGN}$ estimates are not shown.}
 \label{fig:sample_property}
\end{figure}

The simulations are performed using a parameter grid designed to represent the observed properties of IMBH candidates as shown in Figure~\ref{fig:sample_property}, including BH mass ($\log M_{\rm BH}=5.0,5.5,6.0$), $r$-band magnitude ($m_r=16,17,18,19$ mag), AGN flux fraction ($f_{\rm AGN}=0.05,0.1,0.2$), and intrinsic asymptotic SF amplitude (${\rm SF}_{\infty}=0.1,0.2,0.3,0.4,0.5$ mag). The adopted amplitude range is motivated by the long-term variability analysis. 
The selected variable subsample has an observed host-diluted variability amplitude of ${\rm SF}_{\infty}\sim0.028$ mag. Applying an approximate host-dilution correction using its median spectroscopic AGN fraction ($f_{\rm AGN}\sim0.08$) yields an estimated intrinsic variability amplitude of $\sim0.35$ mag. Given the uncertainties in this correction, we adopt a grid bracketing this inferred variability scale, with the nearest value of ${\rm SF}_{\infty}=0.3$ mag used as the fiducial case.
This setup results in 180 parameter combinations. Host-galaxy dilution is included through
\[
F_{\rm total}=F_{\rm AGN}+F_{\rm host},
\]
where $f_{\rm AGN}=F_{\rm AGN}/F_{\rm total}$ suppresses the observed variability amplitude in the small-variation limit.

We simulate two observational scenarios under ZTF-like conditions:
\begin{enumerate}
\item \textbf{ZTF fixed-$t_{\rm span}$}: The monitoring baseline is matched to our observed intranight sample, adopting a median continuous duration of 3.9~hours (0.161~days) and a cadence of 4.4 minutes (53 raw measurements, subsequently binned by a factor of 3 into $\sim$17 epochs). Photometric uncertainties are generated using an empirical power-law relation derived from 12,058 dual-image difference measurements:
\[
\sigma_{\rm ZTF}(m_r)=0.0085\times10^{0.4(m_r-17.0)}+0.0041\ {\rm mag}.
\]
Gaussian noise drawn from $\mathcal{N}(0,\sigma_{\rm ZTF}^{2})$ is added to each simulated magnitude measurement.

\item \textbf{ZTF $\tau$-spanning}: The monitoring baseline is extended to approximately one damping timescale ($t_{\rm baseline}\sim\tau$), distributed over $\lceil\tau/1.0\rceil$ consecutive nights. Each night contains a 3.9-hour monitoring sequence sampled at 4.4-minute cadence, with randomized starting times during the first 8 hours of darkness and separated by $\sim20$-hour daytime gaps.
\end{enumerate}

For each parameter combination, we perform 1000 independent Monte Carlo realizations. Variability recovery is evaluated using the same $F$-test selection criterion as in the observational analysis, adopting a 99\% confidence threshold \citep{Diego2010}. Considering the finite false-positive probability of this threshold ($\alpha\sim0.01$), pure-noise realizations produce an expected detection fraction of approximately 1\%. We therefore adopt a conservative threshold of 1.5\%, slightly above the nominal statistical false-positive level, to identify parameter regimes where intrinsic AGN variability produces a measurable excess above the observational noise floor.

\subsubsection{Detectability of INOV under ZTF-like Monitoring}
\label{sec:simulation_results}
\begin{figure*}
 \centering      
 \includegraphics[width=\textwidth]{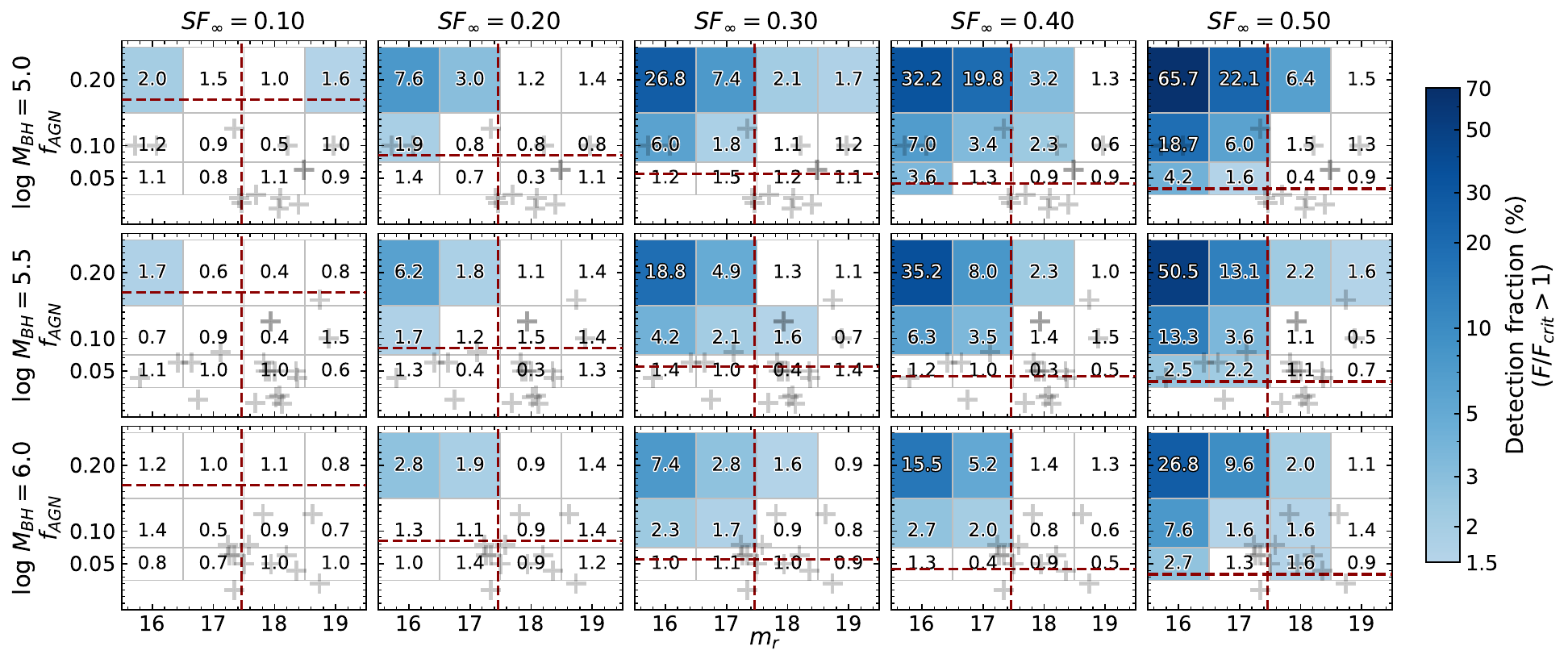}
 \caption{Simulated ZTF INOV recovery fraction under the fixed-$t_{\mathrm{span}}$ scenario (3.9 hr nightly windows). Rows and columns correspond to $\log M_{\rm BH}$ and intrinsic $\mathrm{SF}_{\infty}$, respectively, with each panel showing the recovery fraction in the $m_r$--$f_{\rm AGN}$ parameter space from 1,000 Monte Carlo realizations. The recovery fraction is calculated using the same $F$-test selection criterion adopted for the observational analysis. Regions with recovery fractions below the adopted 1.5\% false-positive threshold are shown without color coding. Red dashed lines indicate the approximate transition boundaries where the host-diluted variability amplitude reaches the observed ZTF photometric noise floor ($\sim0.017$ mag).}
 \label{fig:ztf_recovery_map}
\end{figure*}

Under the ZTF fixed-$t_{\mathrm{span}}$ single-night scenario, the simulated INOV recovery fraction remains low in the parameter regime occupied by our observed IMBH candidates (Figure~\ref{fig:ztf_recovery_map}). Across the full parameter grid, the mean recovery fraction is 3.7\%, while the highest recovery fraction reaches 65.7\% only for the most favorable combinations ($\log M_{\rm BH}=5.0$, $\mathrm{SF}_{\infty}=0.50$~mag, $m_r=16$, and $f_{\rm AGN}=0.20$). At the fiducial variability amplitude (${\rm SF}_{\infty}=0.3$ mag), sources in this observed-candidate regime with $f_{\rm AGN}\lesssim0.1$ and $m_r\gtrsim17$ exhibit recovery fractions of 0.4--2.1\%, with a mean value of $\sim1.2\%$. For the 163 intranight sessions in our sample, the observed-candidate parameter regime therefore corresponds to an expected number of recovered INOV events of $\lesssim2$, consistent with the observed null detection.

The detectability of INOV is primarily governed by the ratio between the host-diluted variability amplitude and the observed photometric noise floor. At the fiducial variability level (${\rm SF}_{\infty}=0.3$ mag), higher recovery fractions are preferentially achieved in regions of parameter space with brighter optical magnitudes and larger AGN contributions, where the intrinsic variability signal remains above the observed ZTF photometric noise floor (red dashed lines in Figure~\ref{fig:ztf_recovery_map}). Most host-dominated and fainter IMBH candidates therefore remain in the noise-limited regime of current ZTF intranight monitoring.

As shown in Figure~\ref{fig:detection_param} (row 1), the recovery fraction is primarily controlled by source brightness, AGN contribution, and intrinsic variability amplitude after host dilution. Increasing $f_{\rm AGN}$ from 0.05 to 0.20 increases the mean recovery fraction from 1.1\% to 7.5\%, while increasing the intrinsic amplitude from $\mathrm{SF}_{\infty}=0.1$ to 0.5~mag increases the recovery fraction from 1.0\% to 7.8\%. In contrast, the dependence on BH mass is relatively weak because the adopted damping timescale increases with BH mass in the simulation model, reducing the variance sampled within a fixed intranight monitoring window. These trends indicate that future high-cadence INOV searches would benefit from targeting brighter sources with larger AGN fractions and stronger expected variability amplitudes.

Together, these results indicate that the null INOV detection in our sample is consistent with the limited recovery efficiency expected for host-dominated IMBH candidates. After seeing-dependent host-galaxy systematics are mitigated through difference-image photometry, current ground-based ZTF-like monitoring remains primarily limited by its sensitivity to low-amplitude, host-diluted variability signals.

\subsubsection{Implications for Future Monitoring Strategies}\label{sec:future_strategy}
We next explore the implications of extended monitoring baselines for variability detection. Beyond the fixed-$t_{\rm span}$ case, the $\tau$-spanning strategy allows us to distinguish whether additional monitoring epochs improve the detectability of individual intranight events or primarily enhance the characterization of longer-timescale stochastic variability. We therefore analyze the simulations in two ways: by treating each nightly session independently and by combining the complete multi-night baseline.

\begin{figure}[t!]
 \centering
 \includegraphics[width=\columnwidth]{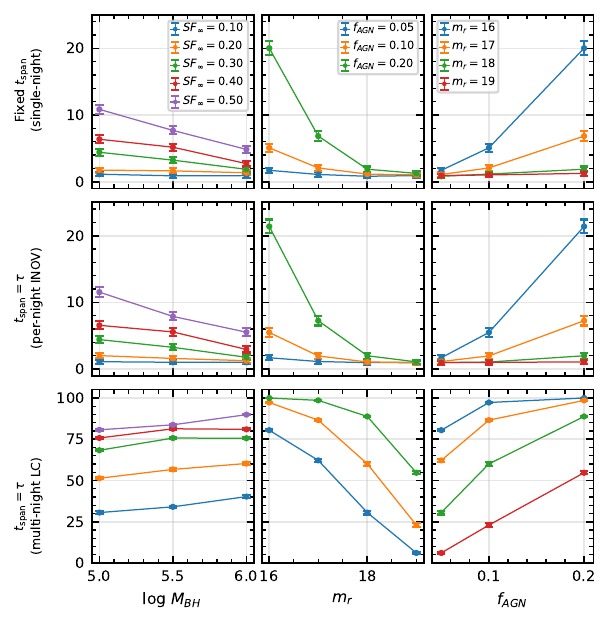}
 \caption{Mean ZTF INOV recovery fractions as a function of individual simulation parameters for three temporal-baseline scenarios: a single-night fixed-$t_{\mathrm{span}}$ window (3.9~hr; Row 1), $\tau$-spanning individual nightly blocks analyzed independently (Row 2), and the complete $\tau$-spanning multi-night baseline analyzed as a single LC (Row 3). Panels show the mean recovery fraction with binomial standard errors from 1,000 Monte Carlo realizations, grouped by $\log M_{\rm BH}$ (left), $m_r$ (middle), and $f_{\rm AGN}$ (right). Curves are color-coded by the corresponding secondary parameter ($\mathrm{SF}_{\infty}$, $f_{\rm AGN}$, and $m_r$, respectively). Rows 1 and 2 exhibit nearly identical per-session recovery fractions, demonstrating that additional short monitoring windows do not increase the probability of individual-night INOV recovery, whereas Row 3 shows the enhanced sensitivity to long-term stochastic variability from an extended baseline.}
 \label{fig:detection_param}
\end{figure}

When individual nights are analyzed independently, the per-session INOV recovery fraction remains nearly unchanged compared with the standard ZTF single-night scenario, with a mean value of 3.8\% compared with 3.7\% for the fixed-$t_{\rm span}$ case (Figure~\ref{fig:detection_param}, row 2). This indicates that extending the monitoring baseline does not improve the per-session recovery probability of INOV, because each monitoring session samples an independent realization of the same short-timescale stochastic variability process. 
Nevertheless, multi-epoch campaigns are expected to increase the cumulative probability of detecting at least one INOV event from a given source, consistent with previous multi-epoch INOV studies (e.g., \citealt{Ojha_etal_2024}).

In contrast to the per-session recovery fraction, combining multiple nights into a continuous $\tau$-spanning baseline substantially improves the recovery of long-term stochastic variability (Figure~\ref{fig:detection_param}, row 3). The mean recovery fraction increases to $\sim66\%$, with clear dependencies on both observational and intrinsic parameters. The recovery fraction decreases from 92.5\% at $m_r=16$ to 28.0\% at $m_r=19$, demonstrating the dominant role of photometric precision. Increasing $f_{\rm AGN}$ from 0.05 to 0.20 raises the recovery fraction from 44.8\% to 85.5\%, while increasing $\mathrm{SF}_{\infty}$ from 0.1 to 0.5~mag increases it from 35.0\% to 84.7\%. These results demonstrate that extended baselines are particularly effective for constraining long-timescale stochastic variability.

To illustrate the impact of improved photometric precision, we repeat the fixed-$t_{\rm span}$ simulations using the expected LSST single-visit photometric uncertainty model \citep{Ivezic_etal_2019}, while keeping the same sampling strategy and intrinsic variability parameters. The LSST photometric uncertainty is calculated using the standard single-visit signal-to-noise relation,
\begin{equation}
\sigma^2(m) = \sigma_{\rm sys}^2 + (0.04-\gamma)\,x + \gamma\,x^2, \qquad x = 10^{0.4(m-m_5)},
\label{eq:lsst_noise}
\end{equation}
with $m_5=24.5$, $\gamma=0.039$, and $\sigma_{\rm sys}=0.005$~mag for the $r$ band. Over the explored magnitude range ($m_r=16$--$19$), the resulting single-visit uncertainties are $\sim0.005$--$0.006$~mag, substantially smaller than those adopted in the ZTF simulations. Under this simplified precision assumption, adopting LSST-like photometric uncertainties increases the mean single-night recovery fraction to 15.7\%, approximately four times higher than the ZTF prediction. The improvement is most significant for intermediate-brightness targets ($m_r=17$--$18$), where the recovery fraction reaches $\sim16$--17\%, compared with $\lesssim3\%$ for ZTF. These results illustrate the potential gain from improved photometric precision alone, while a realistic assessment of LSST performance will additionally require the full survey cadence and observing strategy to be considered.

Finally, we assess the impact of uncertainty in the adopted $M_{\rm BH}$--$\tau$ relation. The empirical relation from \citet{Burke2021} was calibrated primarily using higher-mass BH systems and may underestimate the true damping timescale in the IMBH regime \citep{Su_etal_2024, Zhou_etal_2024, Kim_etal_2026}. A longer true $\tau$ would further reduce the expected variance within a fixed intranight window because the short-timescale DRW variance scales approximately as $\Delta t/\tau$. Therefore, our current single-night recovery estimates should be regarded as conservative. For monitoring strategies designed to probe variability over comparable fractions of the characteristic timescale, the resulting detectability is expected to remain insensitive to the absolute value of $\tau$.

Overall, these simulations highlight the different roles of photometric precision and temporal sampling in constraining variability. Single-session INOV detectability is primarily controlled by photometric precision, AGN contribution, and intrinsic variability amplitude, whereas extended monitoring campaigns mainly enhance the cumulative probability of capturing rare variability events and improve constraints on long-timescale variability.

\subsection{Comparison with Previous INOV Studies}
\label{sec:comparison}
\citet{Gopal-Krishna_etal_2023} reported a high INOV detection fraction of $\sim22\%$ ($8/36$ variable sessions) for a sample of 12 low-mass AGNs monitored with 1-m class telescopes. This differs from our null result (0/163 sessions; $<0.6\%$). However, a direct source-by-source comparison indicates that the two measurements are not statistically inconsistent given the stochastic nature of AGN variability and the different selection functions of the two samples.

Cross-matching their 12 targets with our parent catalog yields 11 common sources. Only two of these have ZTF coverage satisfying our intranight cadence requirements: J085152.63$+$522833.00 ($\log M_{\rm BH}=5.97$) and J073106.87$+$392644.70 ($\log M_{\rm BH}=6.20$). For J085152.63$+$522833.00, both studies report no significant INOV. For J073106.87$+$392644.70—which exceeds our strict $\log M_{\rm BH}<6$ threshold but is included here for comparison—\citet{Gopal-Krishna_etal_2023} reported one INOV event on 2018-11-22, while their adjacent monitoring epochs on 2017-12-29 and 2019-01-03 showed no significant variability. Our independent ZTF session for this target (MJD~58511; 2019-01-28) also exhibits no significant INOV. Our non-detection therefore does not contradict their findings but reflects the transient nature of INOV.

The different INOV detection fractions between the two studies may partly arise from the distinct sample selection criteria and the resulting AGN-to-host contrast distributions. The targets in \citet{Gopal-Krishna_etal_2023} were pre-selected using both radio and X-ray detections, whereas our sample was constructed from optically selected broad H$\alpha$ emitters with sufficient ZTF intranight coverage. Based on consistent spectral decomposition methods, the previous sample has a median $f_{\rm AGN}=0.23$, with $10/11$ sources having $f_{\rm AGN}>0.10$. In contrast, our sample has a median $f_{\rm AGN}\sim0.06$, with only $8/64$ sources having $f_{\rm AGN}>0.10$. These differences should be interpreted with caution, as neither sample represents a complete census of the low-mass BH population and both are subject to selection effects related to sample construction and observational availability. 

The lower AGN flux fractions in our sample are expected to result in stronger host-galaxy dilution, reducing the observable variability amplitude and making low-amplitude INOV signals more difficult to recover with ground-based monitoring. Therefore, our null detection should not be interpreted as evidence that IMBH candidates lack short-timescale variability; instead, it provides a baseline constraint on the detectable INOV fraction in an optically selected IMBH candidate population without prior radio or X-ray selection.

Methodological differences may also contribute to the different INOV detection fractions. By combining difference-image photometry, empirical uncertainty calibration, and comparison-star validation, our analysis provides an additional approach to suppressing host-related and atmospheric systematics, which are particularly important for nearby host-dominated IMBH candidates.

\subsection{Multi-wavelength Constraints on Sample Selection Effects}
\label{sec:radio_xray}
To characterize the multi-wavelength properties of our optically selected IMBH candidates, we cross-matched the sample with major radio and X-ray surveys. We note that individual radio or X-ray detections do not uniquely identify a specific accretion mode or emission mechanism; in particular, radio emission may arise from relativistic jets or other processes such as star formation. A detailed classification of the radio and X-ray emission mechanisms is beyond the scope of this work.

We cross-matched the 64 targets with major radio surveys using search radii scaled to the positional uncertainties of each survey (see Guo et al. 2026, in preparation). In total, 23 sources have radio counterparts in at least one survey. Restricting to GHz-frequency surveys commonly used to trace compact AGN-related radio emission (FIRST, RACS-Mid, RACS-High, and VLASS), 15 sources are detected in at least one band, with 12 detected in multiple surveys.

We further cross-matched the sample with major X-ray catalogs, including XMM-Newton, Chandra, and eROSITA, using instrument-dependent search radii. Seventeen sources have robust X-ray detections in at least one catalog. For eRASS-DR1, which does not cover the full sky region of our sample, 45 targets fall within the survey footprint, among which eight are detected, corresponding to a counterpart fraction of 18\% (8/45).

Across the full sample, seven sources are simultaneously detected in both radio and X-ray bands. These multi-wavelength detected systems remain predominantly host dominated, with a median $f_{\rm AGN}\sim0.06$ among the six sources with reliable measurements, and only one source reaching $f_{\rm AGN}\gtrsim0.1$. Therefore, although our optically selected sample contains radio/X-ray detected systems, most targets remain characterized by low AGN-to-host contrast.

\subsection{Physical Interpretation of INOV Constraints} \label{sec:physical_interpretation}
Rapid optical variability in AGNs can arise from several physical mechanisms, including X-ray reprocessing, intrinsic accretion-disk fluctuations, and jet-related components. Specifically, in the X-ray reprocessing scenario, rapid fluctuations originating from the hot corona are reprocessed by the accretion disk, producing short-term modulation of the optical/UV continuum \citep[e.g.,][]{Czerny_etal_2008, Edelson2019}. Alternatively, intrinsic disk fluctuations may introduce stochastic luminosity variations. While global viscous timescales are generally too long to explain intranight variability \citep{Lyubarskii1997, Li_Cao_2008}, local processes such as transient thermal instabilities or ``hot spots'' \citep{Zhang_Bao_1991, Mangalam_Wiita_1993, Czerny_etal_2008}, localized temperature fluctuations independent of large-scale disk structures \citep{Dexter_Agol_2011}, magnetically coupled disk-corona fluctuations \citep{Sun_etal_2020, Ojha_etal_2024}, and local fluctuations modulated by underlying global variability modes \citep{Cai_etal_2018, Cai_etal_2020} have been proposed as possible drivers of rapid optical variability. In addition, weak or uncollimated jet/outflow components may contribute to micro-variability in some low-mass AGN systems \citep{Czerny_etal_2008}.

However, our null INOV detection does not provide a direct constraint on the relative contribution of these mechanisms, because the expected variability amplitudes after host-galaxy dilution are generally below the sensitivity limit of current ZTF intranight observations. Therefore, the absence of detected INOV primarily reflects limited observational detectability rather than the absence of rapid accretion-related fluctuations in IMBH candidates.

Furthermore, even when INOV is detected, single-band optical monitoring alone cannot uniquely determine the origin of rapid variability in IMBH candidates. Coordinated multi-band observations and inter-band time-delay measurements will be required to distinguish between these variability scenarios and reveal the physical processes governing low-mass accretion systems.
Future high-precision time-domain surveys with improved photometric precision and multi-band coverage, such as the LSST, will provide enhanced sensitivity to low-amplitude variability through improved temporal sampling and photometric accuracy. In addition, space-based facilities such as the China Space Station Telescope (CSST), particularly its Multi-Channel Imager (MCI), with high-resolution multi-band imaging capabilities \citep{CSST_2026}, will provide complementary opportunities to improve the separation of nuclear and host-galaxy emission and investigate the physical origin of rapid variability in low-mass accreting BHs.

\section{Conclusions}\label{sec:conclusion}
From a parent sample of 1,447 broad H$\alpha$-selected low-mass BH candidates, we construct a sample of 64 IMBH candidates with 163 ZTF intranight monitoring sessions. We combine difference-image photometry, variability validation, long-term variability analysis, and Monte Carlo simulations to investigate the presence and detectability of short-timescale variability in IMBH candidates. The main results are summarized as follows:

\begin{enumerate}
\item \textbf{Robust INOV constraint and host-galaxy systematics:} 
Using SExtractor dual-image difference photometry combined with statistical tests, comparison-star validation, and visual inspection, we detect no robust INOV in any of the 163 monitoring sessions, corresponding to an empirical session-level detection fraction of zero. Given the sample size, the minimum non-zero detectable fraction is $1/163\approx0.6\%$. In contrast, conventional ZTF PSF-fit photometry identifies $\sim55$ apparent variable sessions. Further analysis indicates that this discrepancy is primarily associated with seeing-dependent changes in the relative contributions of compact nuclear and extended host components, producing spurious variability that can mimic intrinsic short-timescale variability. This highlights the importance of robust photometric methodologies for reliable INOV measurements in host-dominated IMBH systems.

\item \textbf{Long-term variability beyond intranight timescales:} 
The ensemble SF analysis spanning $\Delta t\sim0.003$--$1600$ days shows that intrinsic variability remains unresolved at intranight timescales within the current photometric precision, but becomes measurable at multi-day lags for the long-term variable subsample. This timescale dependence provides an empirical explanation for the absence of detectable INOV in our sample.

\item \textbf{INOV detectability from simulations:} 
Monte Carlo simulations show that ZTF-like single-night monitoring recovers INOV with a mean probability of $\sim1.2\%$ for the variability amplitudes inferred from the long-term analysis. The recovery probability is mainly controlled by source brightness, AGN fraction, intrinsic variability amplitude, and photometric precision. For the 163 monitoring sessions, this low recovery probability corresponds to an expected number of recovered INOV events of $\lesssim2$, consistent with the observed null detection.
\end{enumerate}

Overall, our results demonstrate that reliable INOV constraints for IMBH candidates require careful control of seeing-dependent systematics and sufficient photometric precision for low-amplitude variability measurements. The absence of detected INOV does not imply the absence of rapid accretion variability, but reflects the limited ability of current observations to recover weak short-timescale signals. Future high-precision, multi-band time-domain surveys with improved spatial resolution and extended temporal sampling will provide stronger constraints on rapid variability and accretion physics in IMBH candidates.

\begin{acknowledgements}
WWZ is supported by the Shanghai Natural Science Foundation Youth Project (Grant No. 25ZR1402546), Strategic Priority Research Program of the Chinese Academy of Sciences (Grant No. XDB0800302) and the National Key Research and Development Program of China (Grant No. 2025YFA1614102). 
WWZ and HXG are supported by the National Key R\&D Program of China (Grant No. 2022YFF0503402, 2023YFA1607903), and Future Network Partner Program, CAS (Grant No. 018GJHZ2022029FN), Overseas Center Platform Projects, CAS (Grant No. 178GJHZ2023184MI). 
LCH was supported by the National Science Foundation of China (12233001) and the China Manned Space Program (Grant No. CMS-CSST-2025-A09).
ACG is partially supported by the CAS ``President's International Fellowship Initiative (PIFI)'' (Grant No. 2026PVA0040).
MFG is supported by the Shanghai Pilot Program for Basic Research-Chinese Academy of Science, Shanghai Branch (Grant NO. JCYJ-SHFY-2021-013), the National SKA Program of China (Grant No. 2022SKA0120102), the science research grants from the China Manned Space Project (Grant No. CMSCSST-2021-A06), and the Original Innovation Program of the Chinese Academy of Sciences 715 (Grant No. E085021002). 
\end{acknowledgements}

\facility{ZTF}
\software{SExtractor \citep{Bertin1996},
          Astropy \citep{Astropy2018},
          NumPy \citep{Harris2020},
          SciPy \citep{Virtanen2020},
          Matplotlib \citep{Hunter2007},
          Photutils \citep{Bradley2020}
         }

\bibliographystyle{aasjournal}
\bibliography{ms.bib}{}

\appendix
\section{Definitions of Variability Amplitude Metrics}
\label{app:var_amp}
For completeness, we define the variability amplitude metrics used to characterize preliminary INOV candidates. The peak-to-peak amplitude is calculated following \citet{Heidt_Wagner1998}:
\begin{equation}
\phi =
\sqrt{(m_{\rm max}-m_{\rm min})^2-2\sigma_m^2},
\end{equation}
where $m_{\rm max}$ and $m_{\rm min}$ are the maximum and minimum magnitudes in the binned LC, and $\sigma_m^2$ is the mean squared photometric uncertainty.

The fractional variability amplitude is calculated following \citet{Vaughan2003}:
\begin{equation}
V=
\frac{1}{\langle f\rangle}
\sqrt{\Delta_f^2-\sigma_f^2},
\end{equation}
where
\begin{equation}
\Delta_f^2=
\frac{1}{N-1}
\sum_{i=1}^{N}(f_i-\langle f\rangle)^2,
\end{equation}
and
\begin{equation}
\sigma_f^2=
\frac{1}{N}
\sum_{i=1}^{N}f_{{\rm err},i}^{2}.
\end{equation}
Here, $f_i$ and $f_{{\rm err},i}$ represent the flux and uncertainty of the $i$-th binned measurement.

\section{Surface Brightness Profile and Structural Measurements}
\label{app:SB_profiles}
For the two representative seeing-induced variability examples, we measure the surface brightness (SB) profiles from $20\arcsec\times20\arcsec$ science-image cutouts centered on each target. The background level is estimated using iterative $3\sigma$ sigma-clipped median subtraction. The SB profiles are extracted in concentric annuli from $0.5\arcsec$ to $6.0\arcsec$ with a radial width of $0.5\arcsec$, and the corresponding encircled flux profiles are obtained from cumulative aperture photometry. These observed profiles are not PSF-deconvolved, as our goal is to characterize the seeing-dependent changes in the observed light distribution rather than recover intrinsic structural parameters.

We characterize the central light concentration using two structural metrics: (1) the concentration index, defined as the fraction of flux enclosed within $1.0\arcsec$ relative to that within a $3.0\arcsec$ aperture; and (2) the peak-to-baseline contrast, defined as $\mathrm{SB}(0$--$0.5\arcsec)/\mathrm{SB}(5$--$5.5\arcsec)$, which quantifies the relative contribution of the central nucleus and extended host component.

\section{Details of Long-term Variability Analysis}
\label{app:longterm_details}
This appendix provides additional details on the construction of the long-term difference-image LCs, the identification of long-term variable candidates, and the ensemble SF analysis described in Section~\ref{sec:longterm}.

\subsection{Adaptive Binning of Long-term Light Curves}
\label{app:longterm_binning}
To account for the non-uniform temporal sampling and different signal-to-noise ratios among targets, we adopt an adaptive target-dependent binning procedure for the long-term difference-image LCs. The bin size for each target is determined by simultaneously satisfying the following criteria:
(1) The final LC contains at least 10 independent temporal bins to ensure sufficient sampling for variability characterization (with a resulting sample median of $\sim 140$ bins).
(2) The minimum bin width is set to three times the median night-to-night cadence, reducing the influence of closely spaced observations.
(3) The fractional variability amplitude $V$ converges with increasing bin size, with the adopted bin size requiring a change of less than 10\% between consecutive trial values.
The convergence criterion is comparable to the median fractional uncertainty of the measured variability amplitudes ($\sim13\%$), ensuring that the adopted binning does not introduce significant systematic changes in the variability measurements. This procedure results in binning timescales of 5--10 days, with a sample median of 7 days.

\subsection{Selection of Long-term Variable Candidates}
\label{app:longterm_selection}
To identify robust long-term variable candidates, we applied multiple variability criteria to the adaptively binned difference-image LCs. The selection was designed to minimize the impact of residual photometric systematics and to identify sources with variability amplitudes significantly exceeding the measurement uncertainties.
For each target, we required the simultaneous satisfaction of three statistical criteria:
(i) an $F$-test significance threshold of $F/F_c>1$;
(ii) a $\geq3\sigma$ detection of the intrinsic fractional variability amplitude $V$; and
(iii) a $\geq3\sigma$ detection of the peak-to-peak variability amplitude $\phi$.

These criteria identify 23 statistically variable candidates among the 64 targets. We then visually inspect their long-term LCs and retain sources showing coherent variability over the full ZTF baseline. Here, ``visually coherent'' refers to variability reflected in the overall LC behavior rather than being driven by a few isolated measurements. Four targets---J080910.72+110619.1, J083021.81+183031.2, J093408.60+175644.0, and J101807.60+011245.0---meet this additional criterion and are adopted as the long-term variable subsample for the subsequent ensemble SF analysis.

This additional selection is motivated by the ensemble SF behavior of the statistically selected candidates. When all 23 candidates are included, the observed variance is comparable to or smaller than the estimated noise variance in most lag bins, and the ensemble SF cannot be measured in these bins. In contrast, the four visually selected candidates exhibit rising ensemble SF profiles, allowing the long-term variability of sources with clearly detected variability to be characterized.

The remaining 60 targets, which do not meet the adopted long-term variability criteria, are used as a variability-undetected reference sample for comparison. This classification is based solely on the absence of detectable variability under the adopted criteria and does not imply that these sources are intrinsically non-variable.

\subsection{Ensemble Structure Function Construction and Uncertainty Estimation}
\label{app:sf_construction}
For each subsample, all unique epoch pairs are generated within individual sources, while cross-source pairs are excluded to avoid introducing artificial correlations. The resulting pairs are grouped into logarithmically spaced time-lag bins. The intranight and long-term measurements are combined following the procedure described in Section~\ref{sec:longterm} to construct a continuous SF over the full temporal baseline.

The uncertainties of the SF amplitudes are estimated using non-parametric bootstrap resampling ($B=1000$) over individual sources rather than individual epoch pairs. This approach preserves correlations introduced by multiple measurements from the same object and provides a more realistic uncertainty estimate for the ensemble SF.

Because the selected long-term variable subsample contains only four or three sources, its bootstrap uncertainties are sensitive to individual objects and should therefore be interpreted cautiously. In comparison, the variability-undetected reference sample contains 60 sources and a substantially larger number of contributing epoch pairs, resulting in more stable ensemble statistics.

The noise-inclusive RMS dispersion of the variability-undetected reference sample is used to characterize the observed photometric noise floor. During subsequent SF model fitting, the lag centers are fixed to the median time separation within each bin. The horizontal uncertainties shown in Figure~\ref{fig:ensemble_SF} represent the adopted lag-bin widths.

\end{CJK}

\end{document}